\documentclass[preprint,12pt]{elsarticle}

\usepackage{amssymb}
\usepackage{amsmath}
\makeatletter
\def\ps@pprintTitle{%
   \let\@oddhead\@empty
   \let\@evenhead\@empty
   \def\@oddfoot{\reset@font\hfil\thepage\hfil} % Keeps page number centered
   \let\@evenfoot\@oddfoot
}
\makeatother

\begin{document}

\begin{frontmatter}



\title{Turbulent mixing of a hydrogen jet in crossflow: direct numerical simulation and model assessment} %% Article title


\author[label1]{Yiqing Wang} %% Author name
\author[label1]{Chao Xu\corref{cor1}} %% Author name
\author[label1]{Riccardo Scarcelli} %% Author name
\author[label2]{Ben Cantrell}
\author[label2]{Jon Anders}
\author[label3]{Sameera Wijeyakulasuriya}

\cortext[cor1]{Corresponding author: chaoxu@anl.gov}

\affiliation[label1]{
            addressline={Argonne National Laboratory}, 
            city={Lemont},
            % postcode={60437}, 
            state={IL 60437},
            country={USA}}
            
\affiliation[label2]{
            addressline={Caterpillar Inc.}, 
            city={Peoria},
            % postcode={61614}, 
            state={IL 61614},
            country={USA}}
            
\affiliation[label3]{
            addressline={Convergent Science, Inc.}, 
            city={Madison},
            % postcode={53705}, 
            state={WI 53705},
            country={USA}}            
%% Abstract
\begin{abstract}
A numerical study for a hydrogen (H$_2$) jet in an air crossflow (JICF) was performed using direct numerical simulation (DNS), large eddy simulation (LES),  and Reynolds-averaged Navier–Stokes (RANS) approaches, based on a geometry representative of key aspects of port fuel injection (PFI) in a H$_2$-fueled heavy-duty internal combustion engine. The focus was placed on the H$_2$ mixing process and the turbulent species flux model used in the latter two approaches. Based on the DNS data, the performance of LES and RANS on predicting the turbulent flow fields and mixing process was comprehensively evaluated. Results showed that LES performs very well in predicting both the mean velocity and the Reynolds stress. In contrast, RANS significantly under-predicts all Reynolds stress components, while predicting the mean flow field relatively well. Regarding the H$_2$ mixing prediction, LES shows an excellent agreement with DNS, while RANS significantly under-predicts the mixing process. The underlying reasons for the poor performance of RANS were identified by extracting turbulent transport properties used in RANS approach from DNS data. It was found that the turbulent diffusivity used in RANS is much smaller than that derived from DNS, which is attributed to the over-prediction on turbulent Schmidt number ($Sc_t$), as well as the under-prediction on turbulent viscosity. By further analyzing the anisotropic components of $Sc_t$ and the misalignment angle between turbulent species fluxes directly obtained from DNS and those predicted by the RANS mixing model, the commonly used assumption of isotropic turbulent diffusivity in RANS was demonstrated to be invalid for the present configuration. This study provided a unique DNS dataset for H$_2$ jet in a crossflow relevant to H$_2$ PFI engines and generated new insights on improved modeling of turbulent mixing.

\end{abstract}

% \includegraphics[width=1.0\linewidth]{figures/fig_YH2_horizontal.jpg}

% \begin{highlights}
% \item A high-fidelity direct numerical simulation 
% (DNS) was performed for the hydrogen jet in an air crossflow.
% \item The performance of large eddy simulation (LES) and Reynolds-averaged Navier–Stokes (RANS) approaches on predicting the turbulent flow fields and mixing process was comprehensively evaluated against the DNS results.
% \item LES shows excellent agreement with DNS in terms of hydrogen mixing prediction, while RANS significantly under-predicted the mixing process. 
% \item By extracting turbulent transport properties used in RANS from DNS data, it was found that the turbulent diffusivity is under-predicted by RANS, and the assumption of isotropic turbulent diffusivity is not valid.
% \end{highlights}

%% Keywords
\begin{keyword}
Hydrogen injection \sep Hydrogen mixing \sep Turbulent species flux \sep DNS \sep LES \sep RANS



\end{keyword}

\end{frontmatter}



\section{Introduction}
\label{sec1}
Hydrogen (H$_2$) has been considered a promising solution to improve the performance and reduce emissions  of a range of energy and propulsion systems, such as off-road internal combustion engines (ICEs) and industrial gas turbines \cite{2020Akal,2023Hwang,2023Manigandan}. In combustion systems such as H$_2$ ICEs with Port Fuel Injection (PFI), hydrogen is injected into and mixed with air through a jet in crossflow (JICF) configuration. The mixing process is key to achieving the desired level of fuel-air premixing which can strongly impact engine performance and emissions characteristics. Hence, it is important to better understand the mixing process of H$_2$ JICF systems, which would enable improved injector and engine designs toward next-generation hydrogen engines. 

Computational fluid dynamics (CFD) provides a unique tool for characterizing fuel-air mixing processes in JICFs. 
% In the past few decades, computational fluid dynamics (CFD) has become an integral part of the engine design and optimization processes. Hence it is important to better understand the mixing process in H$_2$ JICF and develop accurate mixing models to be used with CFD. This will enable the development of next-generation engines with higher efficiency and lower emissions. 
Previous numerical studies on JICFs have demonstrated that both direct numerical simulations (DNS) which resolve all turbulent scales, and large eddy simulations (LES) which resolve large turbulent scales with sufficiently small grid size, are able to predict the key physical structures in JICF and scalar fields accurately \cite{2007Muppidi,2008Muppidi,2013Bodart,2017Zhang,2017Ryan,2019Uyanwaththa}. For example, Muppidi et al. \cite{2007Muppidi,2008Muppidi} performed DNS studies for a transverse jet in crossflow, and they showed that DNS predictions agreed well with experimental measurement of the mean velocity field, turbulent intensities, and scalar fields. Bodart et al. \cite{2013Bodart} conducted an LES study for an inclined jet in crossflow and showed good agreement between their LES results and experimental data on velocity and scalar fields. However, from a practical standpoint, both LES and particularly DNS are still computationally too expensive to be applied directly to complex industrial problems. Therefore, Reynolds-averaged Navier-Stokes (RANS) based approaches, due to their low computational cost, remain the predominant model used in the iterative engine design process.

In RANS framework, an additional model for the turbulent scalar fluxes is required to close the scalar transport equations, which will significantly influence the accuracy of prediction for the mixing process. Among various existing turbulent scalar flux models, the simplest and most commonly used is the Gradient Diffusion Hypothesis (GDH), which models the turbulent scalar fluxes as
\begin{equation}
\overline{u_i^{'}C^{'}} = - D_t \frac{\partial \overline{C}}{\partial x_i}
\label{eq_gdh_intro}
\end{equation}
where $u_i$ is the $i^{th}$ component of velocity, over-bar represents the time-averaging operation, and $C$ is an arbitrary scalar of interest such as species mass fraction or temperature. The turbulent diffusivity $D_t$ is usually  calculated through the Reynolds analogy with a fixed turbulent Schmidt number $Sc_t$ (or Prandtl number $Pr_t$ if $C$ is temperature) as
\begin{equation}
D_t = \nu_t/Sc_t ,
\label{eq_Dt_intro}
\end{equation}
where $\nu_t$ is the eddy viscosity. Clearly, the accuracy of the mixing model relies on two terms, $\nu_t$ and $Sc_t$. Typically, $Sc_t$ is set in the range of $0.7 \leq Sc_t \leq 1.0$, while in many cases, $Sc_t$ is considered a tunable model constant that is determined case by case.  However, many studies \cite{1993Alvarez,1999He,2010Galeazzo,2011Ivanova} have found that this typical range of $Sc_t$ tends to under-predict the mixing process in JICF problems, and lower values of $Sc_t$ in the range of 0.2-0.3 were recommended by trial and error to achieve better agreement with experimental data. 
%Nevertheless, this does not indicate the actual value of $Sc_t$. 
In addition, Ivanova et al. \cite{2011Ivanova} found that the RANS model also under-predicted the turbulent kinetic energy in JICF, indicating a possible under-prediction of turbulent viscosity $\nu_t$. In this context, the extremely low $Sc_t$ recommended (i.e., $Sc_t=0.2-0.3$) might be just an artifact associated with the under-prediction in $\nu_t$. In particular, Ivanova et al. \cite{2013Ivanova} found that the optimal $Sc_t$ value depends on the RANS turbulence model used. When employing the shear-stress transport model, which under-predicted the turbulence level, the optimal $Sc_t$ was 0.2. In constrast, when a curvature correction modification was applied to properly reproduce the turbulence, an optimal value of 0.5 was found. More interestingly, this optimal $Sc_t$ of 0.5 matches the $Sc_t$ derived from their LES results. A similar optimal $Sc_t$ of 0.6 was also reported in \cite{2016Ling} by extracting $Sc_t$ from LES results conducted by Bodart el al. \cite{2013Bodart}. However, a smaller optimal $Sc_t$ of 0.35 was observed in the LES study conducted by Ryan et al. \cite{2017Ryan} for a skewed jet in crossflow. It is therefore evident that when using GDH, accurate descriptions of both the turbulence model and the Schmidt number are critical. 

Another key simplification in GDH is that it assumes an isotropic turbulent diffusivity, which was also believed to cause errors in mixing prediction for inhomogeneous flows \cite{2002Lakehal,2008Liu}. To allow for anisotropy of $D_t$, Daly and Harlow \cite{1970Daly} proposed
the Generalized Gradient Diffusion Hypothesis (GGDH) by including Reynolds stress components in the definition of $D_t$. Abe and Suga \cite{2001Abe} further proposed the Higher Order Generalized Gradient Diffusion Hypothesis (HOGGDH) to better predict the streamwise component in fully-developed turbulent channel flows by using quadratic products of the Reynolds stresses. Ling et al. \cite{2016Ling} evaluated the performance of GDH, GGDH, and HOGGDH in the RANS calculation with the velocity field provided by an LES simulation for an inclined jet in crossflow. They found that switching from GDH to higher-order models would improve scalar concentration predictions to certain extent, but the effect was significantly smaller compared to the tuning of the model parameters. Similar observations were made in the LES study by Ryan et al. \cite{2017Ryan} which highlighted the importance of model constant optimization.

In all above-mentioned JICF studies, the density ratio between the jet and crossflow was set to unity with constant molecular transport properties and the mixing process was simply treated as a passive scalar transport problem. However, for practical systems with H$_2$ JICF, the density ratio is much smaller than one as H$_2$ is much lighter than air. In addition, all the physical properties, such as density, viscosity and diffusivity, are functions of the local H$_2$ concentration. Besides, the high diffusivity of H$_2$ can lead to a very small molecular Schmidt number in the H$_2$-lean regions, which may further impact the turbulent mixing process. Nevertheless, to the best of our knowledge, currently there is no numerical study particularly targeting on the mixing process within the H$_2$ JICF configuration that is directly relevant to practical engine applications despite the vital importance and the unique characteristics of H$_2$. 

Hence, the objectives of this paper are twofold. First is to generate a high-resolution, fully-resolved DNS data set for a H$_2$ jet in air crossflow configuration that is representative of key aspects of the intake system in a realistic H$_2$-fueled off-road PFI engine. Second is to use the DNS data to assess the turbulence and mixing models in RANS and LES approaches that are widely used in industry for engine CFD. 

%% Labels are used to cross-reference an item using \ref command.

\section{Methodology}
\subsection{Physical problem: a H$_2$ jet in air crossflow}
\begin{figure}[tbp]%% placement specifier
\centering%% For centre alignment of image.
\includegraphics[width=1.0\linewidth]{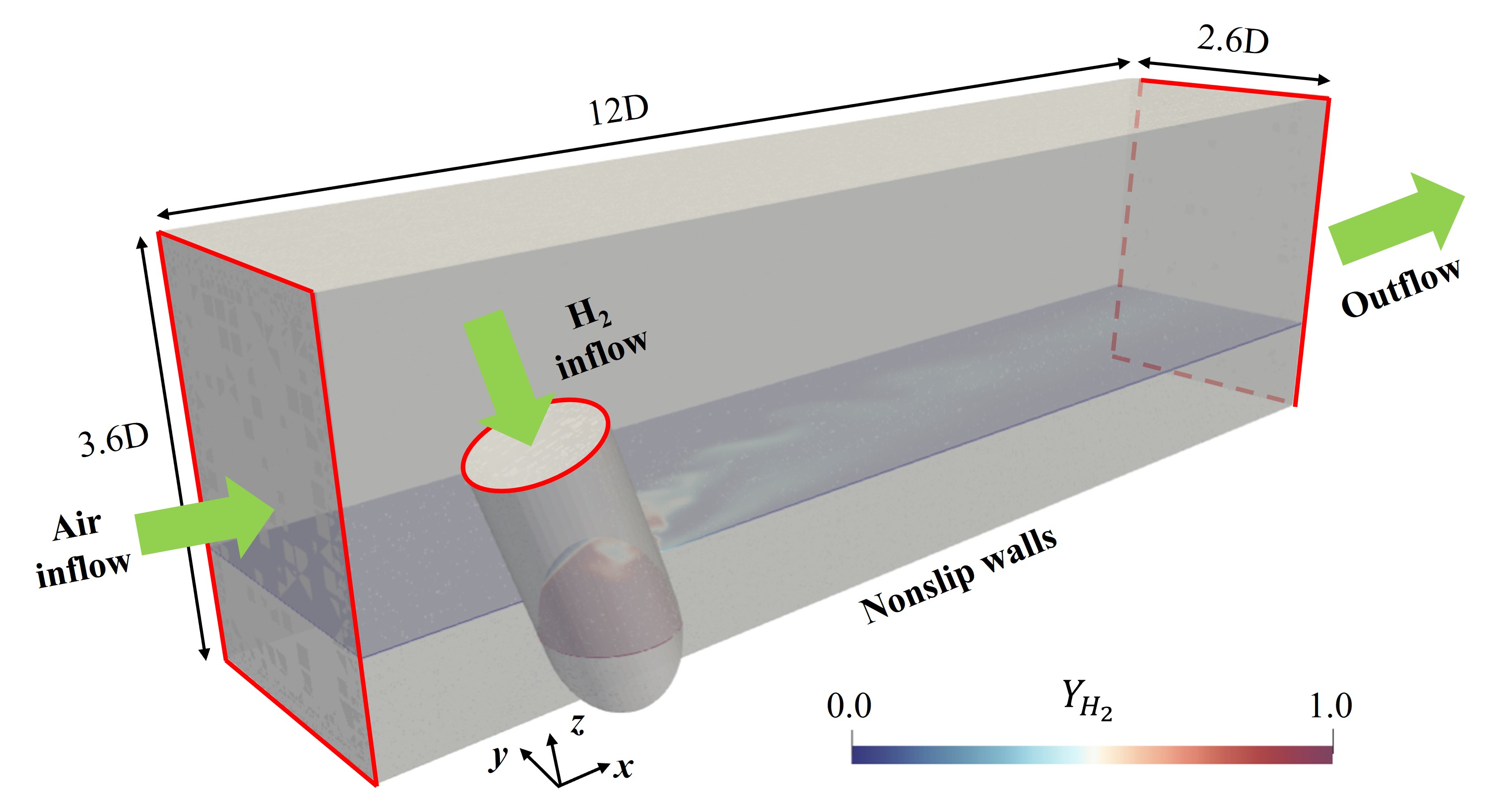}
\caption{Schematic for the flow configuration, computation domain, and boundary conditions. The distribution of H$_2$ mass fraction in a specific horizontal plane cutting through the center of H$_2$ jet hole is also presented for reference.}\label{fig_3dconfig}
\end{figure}

\begin{figure}[tbp]%% placement specifier
\centering%% For centre alignment of image.
\includegraphics[width=1.0\linewidth]{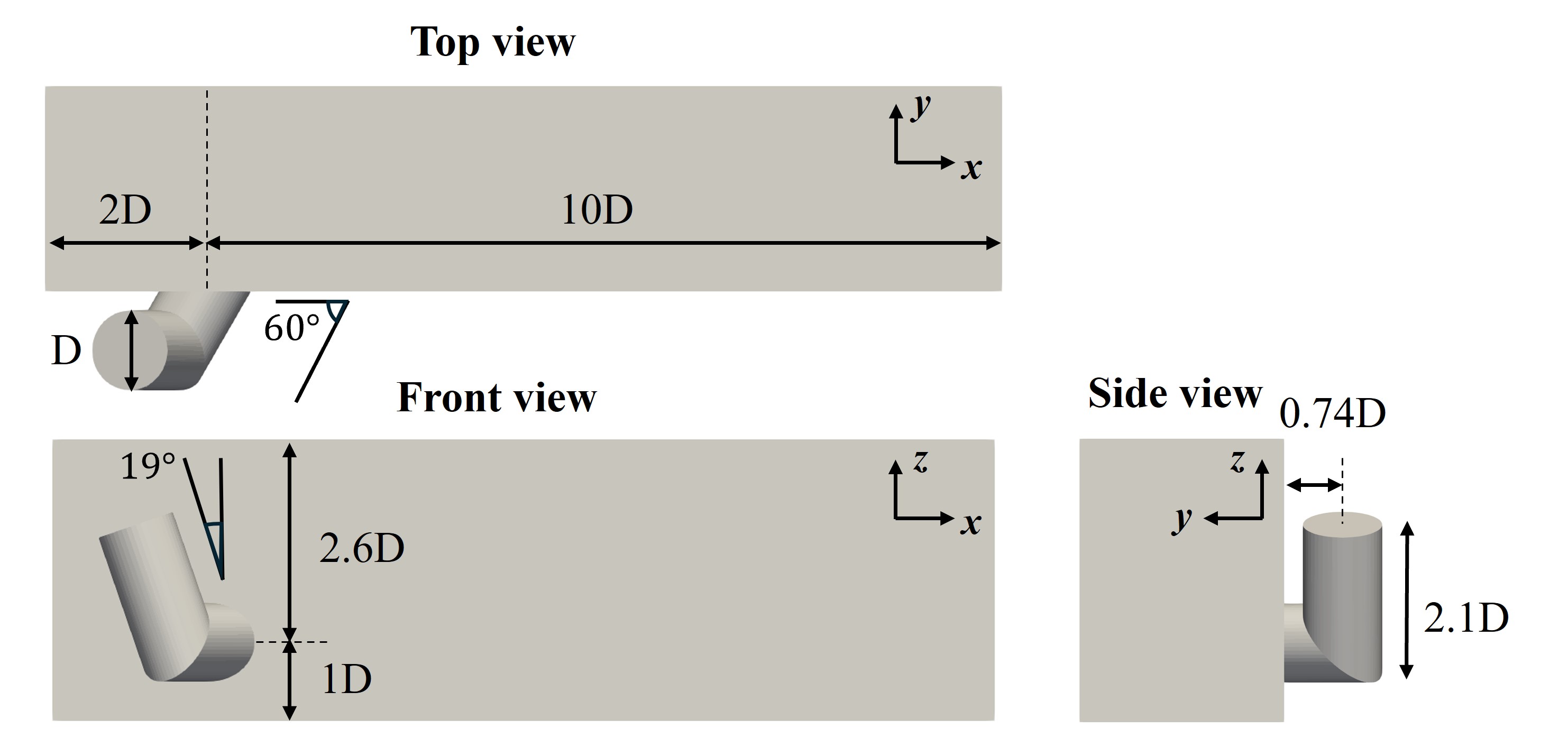}
\caption{Three views of the H$_2$ jet tube and air channel.}\label{fig_3views}
\end{figure}

A H$_2$ jet in air crossflow configuration is considered in the present study. As shown in Figs. \ref{fig_3dconfig} and \ref{fig_3views}, gaseous H$_2$ is injected into air crossflow with rectangular cross-section from the side wall through a round jet inclined 60$^{\circ}$ with respect to the wall. The jet diameter is $D$=19 mm, and the dimension of the crossflow channel is 12$D$ (streamwise, $x$ direction) $\times$ 2.6$D$ (transverse, $y$ direction) $\times$ 3.6$D$ (spanwise, $z$ direction). The center of H$_2$ injection hole is located 2$D$ away from the channel inlet and 1$D$ away from the bottom wall. The temperature of H$_2$ jet and air crossflow inlets are both 325 K. The pressure is 4.3 bar throughout the domain. The bulk velocities of the crossflow $U_{bulk}$ and jet $U_{jet}$ are selected so that Reynolds numbers for the jet and crossflow inlets (based on the hydraulic diameter) are 4600 and 22100, respectively. More details about the layout of H$_2$ jet tube can be found in Fig.~\ref{fig_3views}. Note that the origin of coordinates is placed on the center of jet exit. This configuration is designed to maintain the same relative geometry and flow characteristics as in the real intake port of a heavy-duty H$_2$ ICE, while the jet and crossflow inlet velocities are reduced to allow for a fully-resolved DNS. 

\subsection{DNS configuration}
DNS of the H$_2$ JICF is conducted using the spectral element CFD code Nek5000 \cite{NEK5000}, developed at Argonne National Laboratory. Nek5000 is high-order by design, and highly scalable on modern computing machines. It has been applied to a wide range of fluid dynamic problems \cite{2016Schmitt,2017Giannakopoulos}, including DNS of various wall-bounded flows \cite{2013Khoury,2016Hosseini}, wall-resolved LES of film cooling flows \cite{2022Nunno}, and DNS of turbulent reacting flows \cite{2022Xu}. 
% Therefore, it is believed that the H$_2$ JICF can be accurately simulated by this code even without validation against experiment data (which is not available currently). 

The computational domain is discretized by a non-uniform mesh consisting of 268,000 spectral elements. To ensure all turbulent length scales are well resolved, 7th-order polynomials are used in each element (namely, eight grid points along each spatial direction), resulting in approximately 93 million total grid points. Local refinement is applied to all walls so that the non-dimensional wall-normal distance satisfies $y^+<$1 at all grid points adjacent to the wall. Time integration is based on the semi-implicit second-order backward differentiation method with an operator-integrator factor scheme (OIFS) \cite{2017Patel}, allowing variable time-stepping with a CFL number up to 2. Constant velocity inflow that is normal to the surface is applied to the H$_2$ jet and air channel inlet boundaries with prescribed values, while a non-slip boundary condition with fixed temperature of 325 K is applied to all walls. The time-averaging and Farve-averaging statistics are collected for 10 flow-through times (FTTs) after the flow has reached quasi-statistically steady state. FTT is defined as the time required for the injected H$_2$ to travel from the inlet to the outlet of the channel.
%i.e., 12$D/U_{bulk}$. 
For each FFT, it takes approximately 36 hours when using 4096 CPUs (AMD EPYC 7713 64-Core 2.0 GHz Processor). 

\subsection{LES and RANS configurations}
The same problem is modeled in LES and RANS using CONVERGE CFD code Version 4.0 \cite{Converge}, which is a second-order finite-volume solver with the capability of simulating both incompressible and compressible flows in complex geometries. CONVERGE relies on a base Cartesian grid and a cut-cell approach to generate a computational grid at runtime, and features adaptive mesh refinement (AMR) and fixed embedding (FE) for local mesh refinement. For RANS simulation, the renormalization group (RNG) $k-\epsilon$ model \cite{1986Yakhot} is used, while for LES, the Dynamic Structure model \cite{2002Pomraning} is chosen. Both turbulence models have been widely used in engine CFD simulations \cite{2021Yosri,2023Yosri,2023Wu}.

Similar boundary conditions as in DNS have been used for both LES and RANS. Particularly, the same inlet velocity as in DNS but with a small turbulence intensity (0.1\%) is applied to the two inlet boundaries. Regarding the wall treatments, the Werner and Wengle wall model \cite{1993Werner} is employed for LES, while the scalable wall function \cite{Converge} is used for RANS.

In both LES and RANS, a base grid size of 4 mm is used, with 1 mm cells at walls.  AMR is applied to the entire domain to resolve local flow and mixing fields with refinement criteria based on the local sub-grid velocity of 0.02 m/s and sub-grid H$_2$ mass fraction of 0.003. For RANS, the refinement levels of FE and AMR are both three (denoted as ``Level=3''), corresponding to a minimum cell size of 0.5 mm, which is the typical mesh resolution used for engine simulations. For LES, a finer mesh is used with the levels of FE and AMR being both four (denoted by ``Level=4"), leading to a minimum cell size of 0.25 mm. The resulting $y^+$ is approximately 10 and 5 for RANS and LES, respectively. 

For LES, the total cell count is 8.9 millions and each simulation takes 131 hours for each FFT when using 432 CPUs. For RANS, the total cell count is 1.2 millions and each simulations takes 16 hours for each FFT when using 216 CPUs. The turbulent statistics are collected for 5 FTTs after the flow has reached the statistically steady state.

\subsection{Grid convergence in LES and RANS}
\begin{figure}[tbp]%% placement specifier
\centering%% For centre alignment of image.
\includegraphics[width=0.6\linewidth]{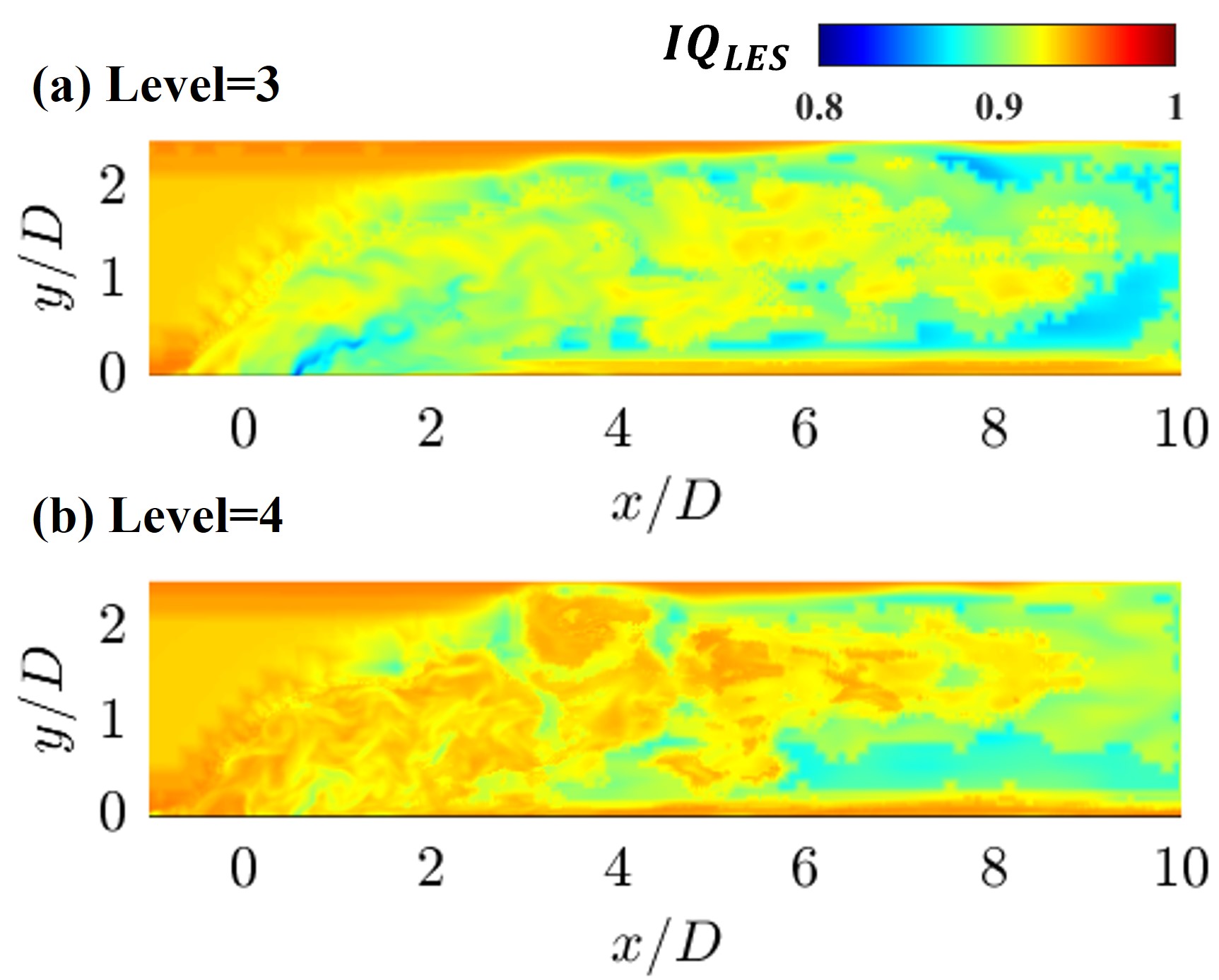}
\caption{Instantaneous distribution of $LES_{IQ}$ for two LES mesh resolutions at the horizontal plane with $z/D$=1.}\label{fig_LES_QI}
\end{figure}

\begin{figure}[tbp]%% placement specifier
\centering%% For centre alignment of image.
\includegraphics[width=0.6\linewidth]{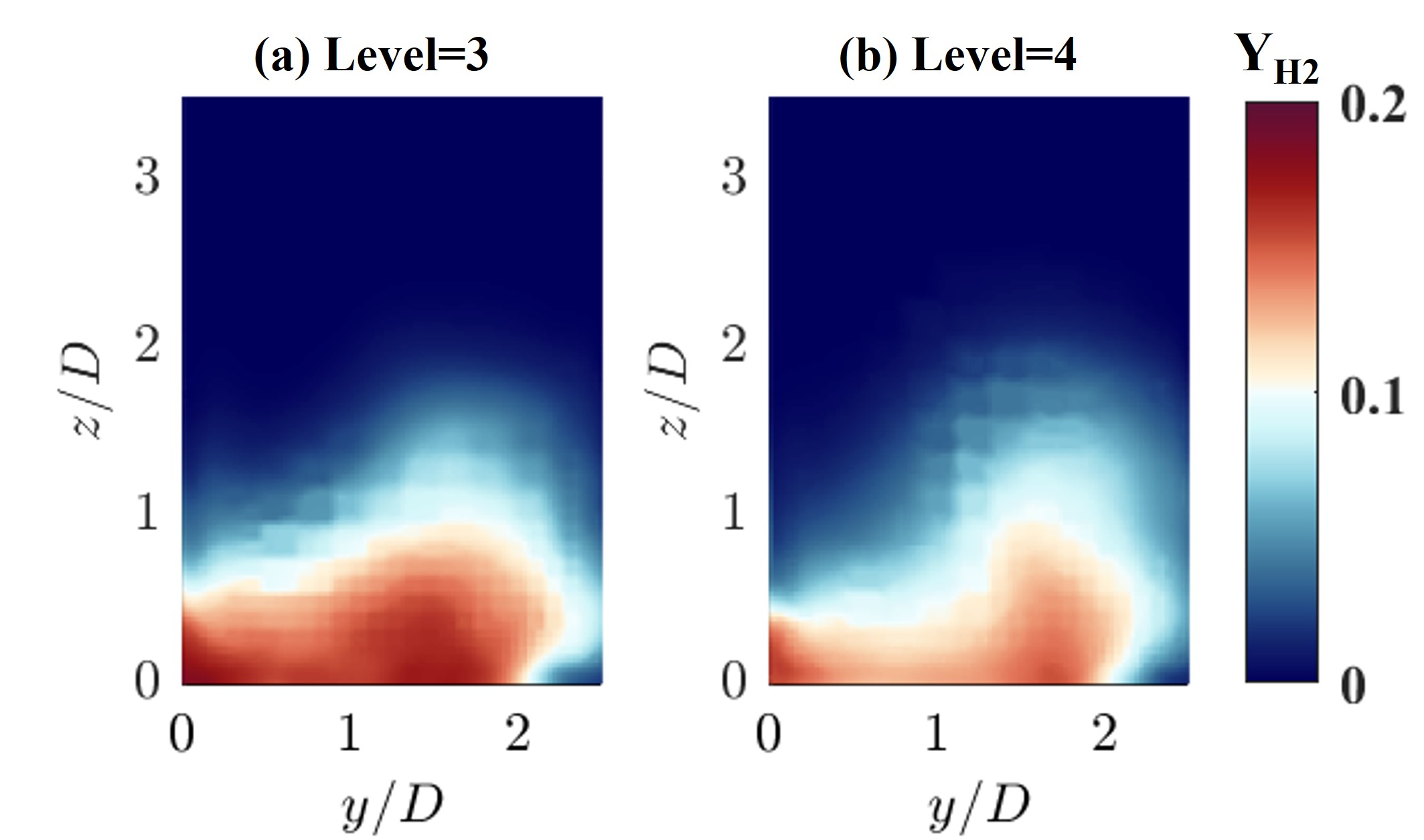}
\caption{Time-averaged H$_2$ mass fraction distributions at the outlet vertical plane $x/D$=10 for two LES mesh resolutions.}\label{fig_LES_YH2}
\end{figure}

To assess the mesh quality for LES, an index of quality ($IQ_{LES}$) proposed by Celik et al. \cite{2005Celik} is computed for two mesh resolutions (Level=3 and Level=4) as follows:
\begin{equation}
    IQ_{LES}=\frac{1}{1+\alpha_\nu(\frac{\nu_{t,eff}}{\nu})^n}
\end{equation}
where $\alpha_\nu$ is a constant of 0.05 and $n$=0.53. $\nu_{t,eff}$ is the effective viscosity which is the sum of the molecular viscosity ($\nu$) and the eddy viscosity ($\nu_t$) \cite{2024Lien}. Figure~\ref{fig_LES_QI} shows the instantaneous distributions of $IQ_{LES}$ at the horizontal plane of $z/D$=1 for these two meshes. As suggested by Celik et al. \cite{2005Celik}, $IQ_{LES}$ of 0.75 to 0.85 can be considered adequate for most engineering applications. In this sense, both meshes are sufficient for the current configuration since $IQ_{LES}$ is higher than 0.85 in most regions for both meshes. The Level=4 mesh indeed shows higher quality than the Level=3 mesh, especially in the region near the H$_2$ jet injection. This results in substantial differences in the distribution of time-averaged H$_2$ mass fraction at the outlet vertical plane between the two meshes as shown in Fig.~\ref{fig_LES_YH2}. It is therefore suggested that the Level=3 mesh is not sufficient to achieve accurate prediction on the mixing process in LES, and thus the Level=4 mesh is used. 

\begin{figure}[tbp]%% placement specifier
\centering%% For centre alignment of image.
\includegraphics[width=0.6\linewidth]{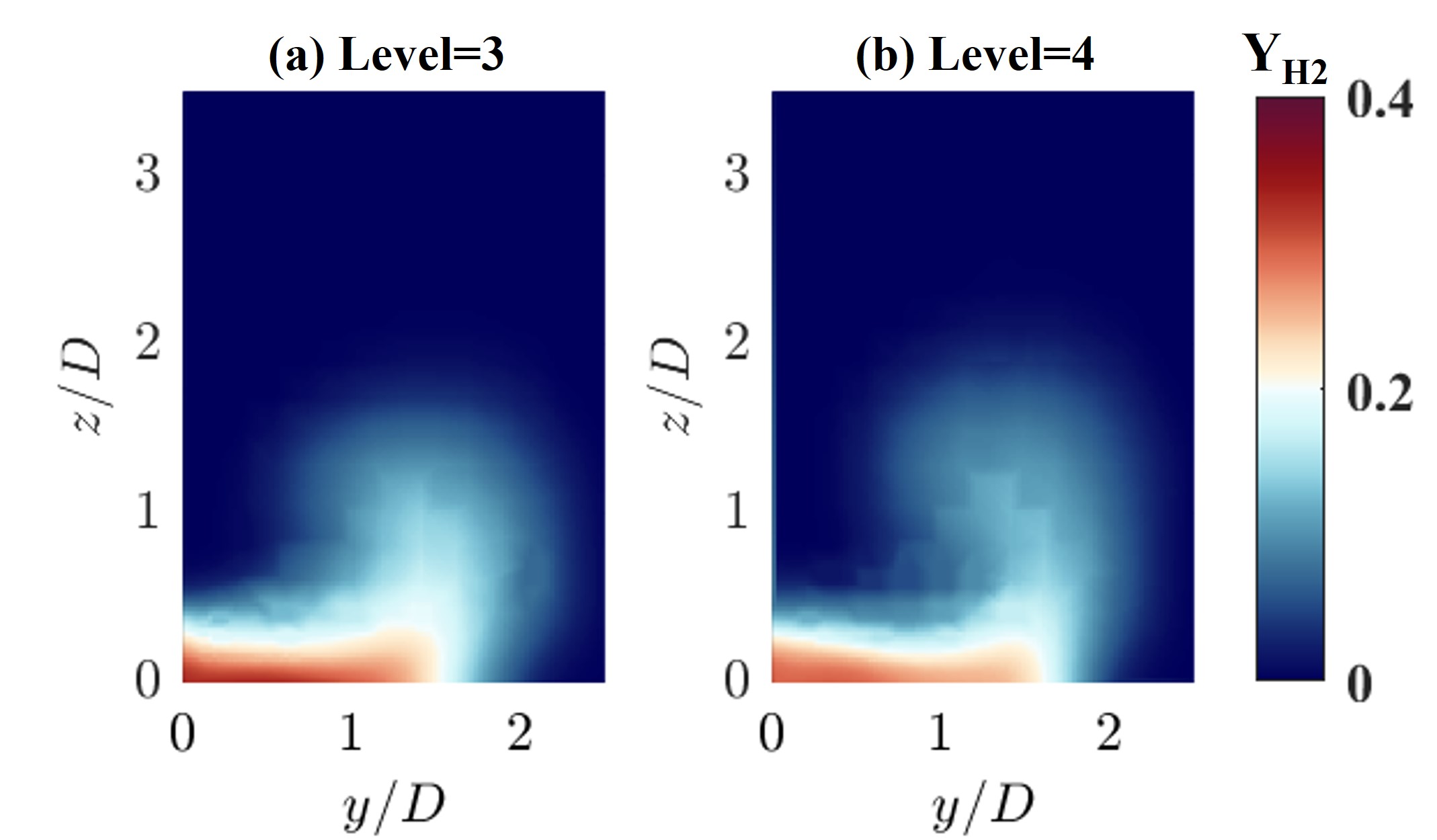}
\caption{Time-averaged H$_2$ mass fraction distributions at the outlet vertical plane $x/D$=10 for two RANS mesh resolutions.}\label{fig_RANS_YH2}
\end{figure}

Unlike LES, RANS prediction in time-averaged H$_2$ mass fraction at the outlet vertical plane is relatively insensitive to the mesh resolution as shown in Fig. \ref{fig_RANS_YH2}, suggesting grid convergence at Level=3. Considering Level=3 mesh is much cheaper for the purpose of practical engine simulations, the Level=3 mesh is used for RANS.

\section{Results and Discussion}

\subsection{Comparison between DNS, LES, and RANS}
In this section, numerical results from DNS, LES and RANS are compared in terms of both flow and H$_2$ mass fractions fields.  DNS serves as the benchmark for assessing the performance of LES and RANS models in predicting the flow dynamics and H$_2$ mixing process.

\subsubsection{Flow fields predictions}
\begin{figure}[tbp]%% placement specifier
\centering%% For centre alignment of image.
\includegraphics[width=1.0\linewidth]{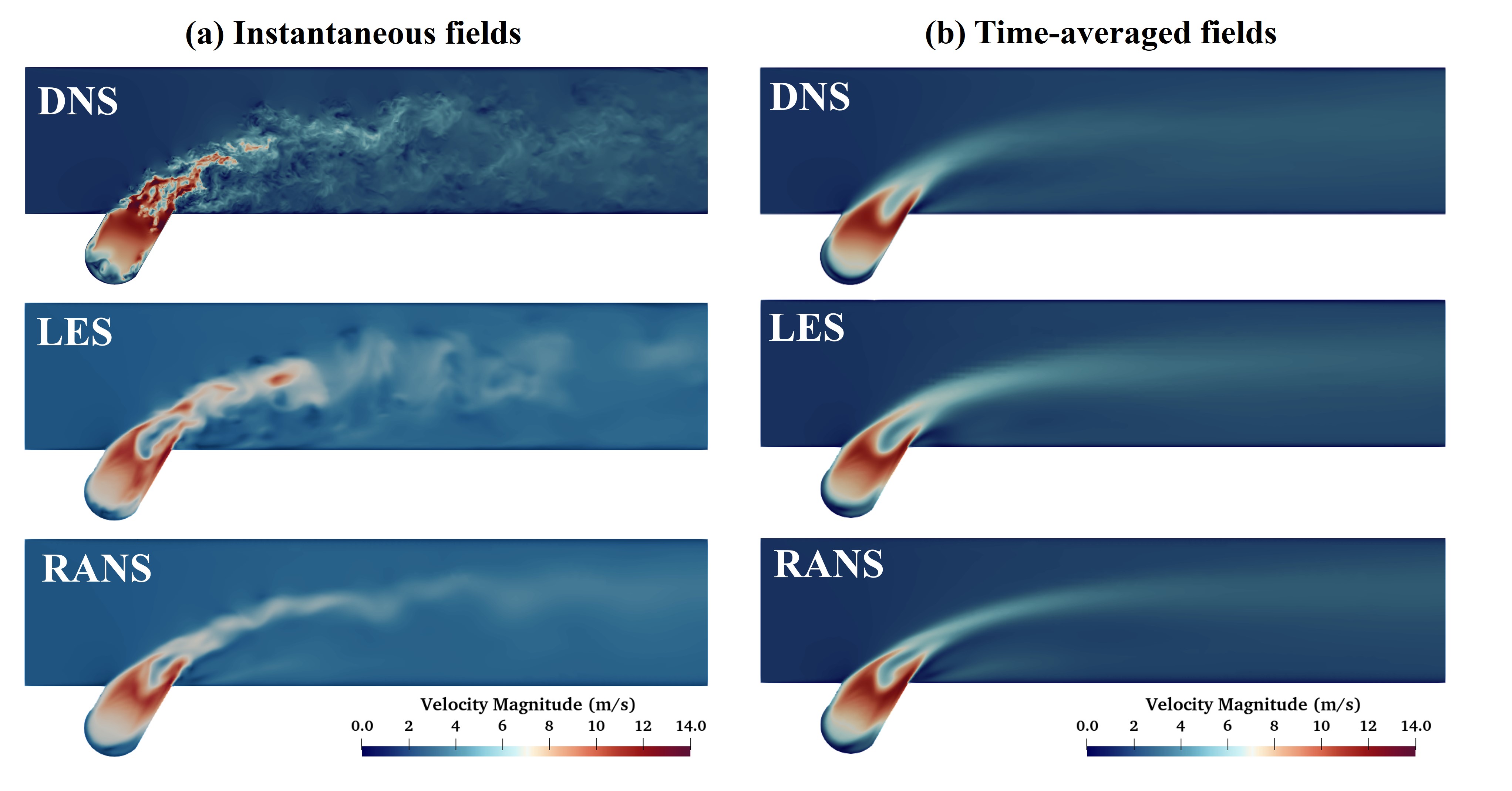}
\caption{Instantaneous (left) and time-averaged (right) velocity magnitude distributions at the horizontal plane with $z/D$=1, predicted by DNS, LES, and RANS.}\label{fig_flow_horizontal}
\end{figure}

Figure~\ref{fig_flow_horizontal} shows instantaneous and time-averaged distributions of the velocity magnitude at the horizontal plane with $z/D$=1, predicted by DNS, LES, and RANS. From the instantaneous field, DNS shows strong turbulence generated by the shear layer between H$_2$ jet and air crossflow, resulting in a large amount of fine flow structure downstream of the jet. LES, as expected, captures the large structure in the flow field. In contrast, RANS only predicts the main flow pattern near the jet exit without any fine flow structure being resolved. For time-averaged flow field, the results from the three numerical approaches are similar in the region close to the jet exit. This is reasonable since all small structures have been filtered out by the averaging operation.

%Nevertheless, the mean jet core downstream of the exit predicted by RANS is less diffused compared with DNS and LES. 

\begin{figure}[tbp]%% placement specifier
\centering%% For centre alignment of image.
\includegraphics[width=0.9\linewidth]{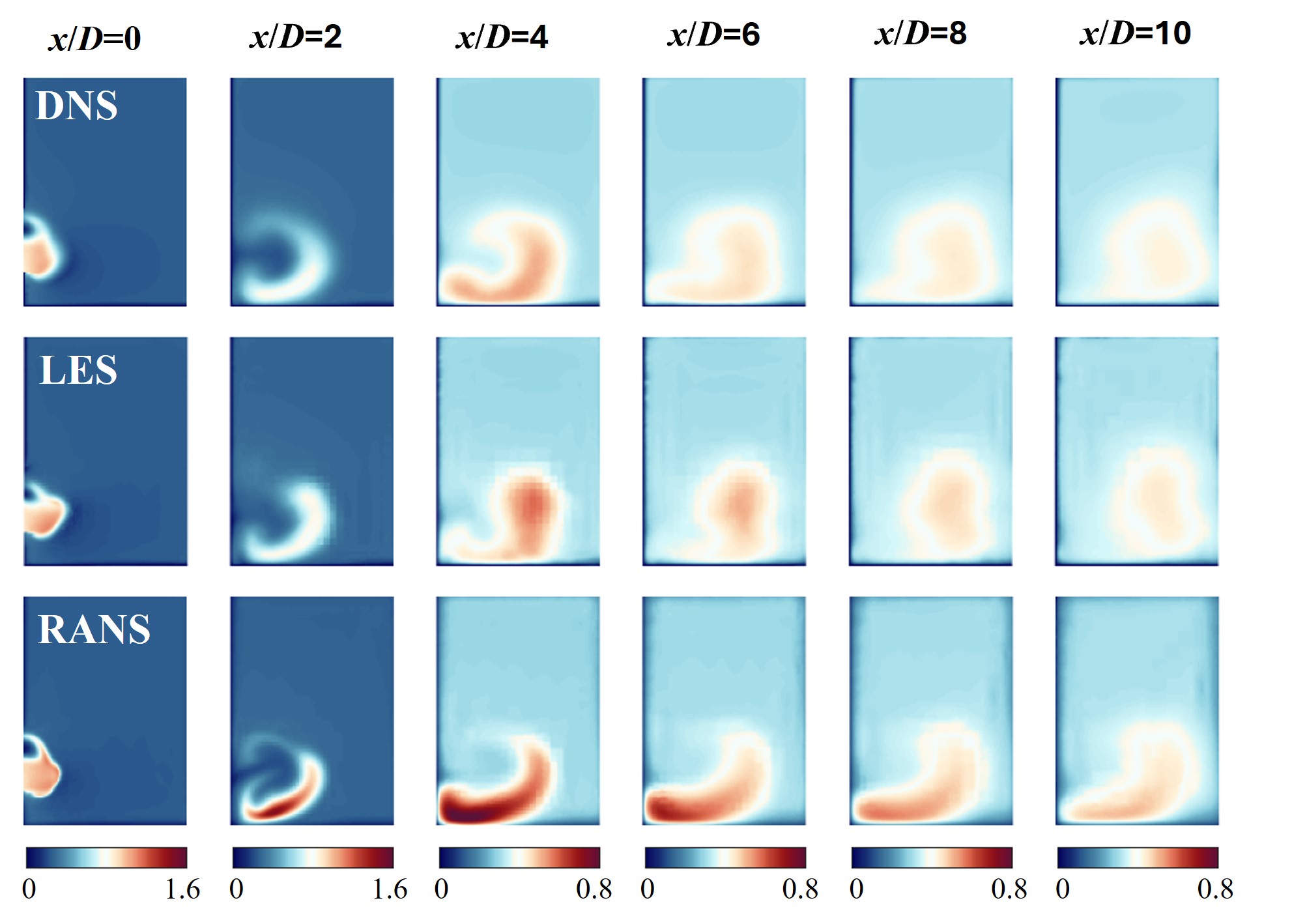}
\caption{Time-averaged velocity component $\overline{u}/U_{jet}$ at different vertical planes with $x/D$=0, 2, 4, 6, 8, and 10, predicted by DNS, LES, and RANS.}\label{fig_vertical_u}
\end{figure}

\begin{figure}[tbp]%% placement specifier
\centering%% For centre alignment of image.
\includegraphics[width=0.8\linewidth]{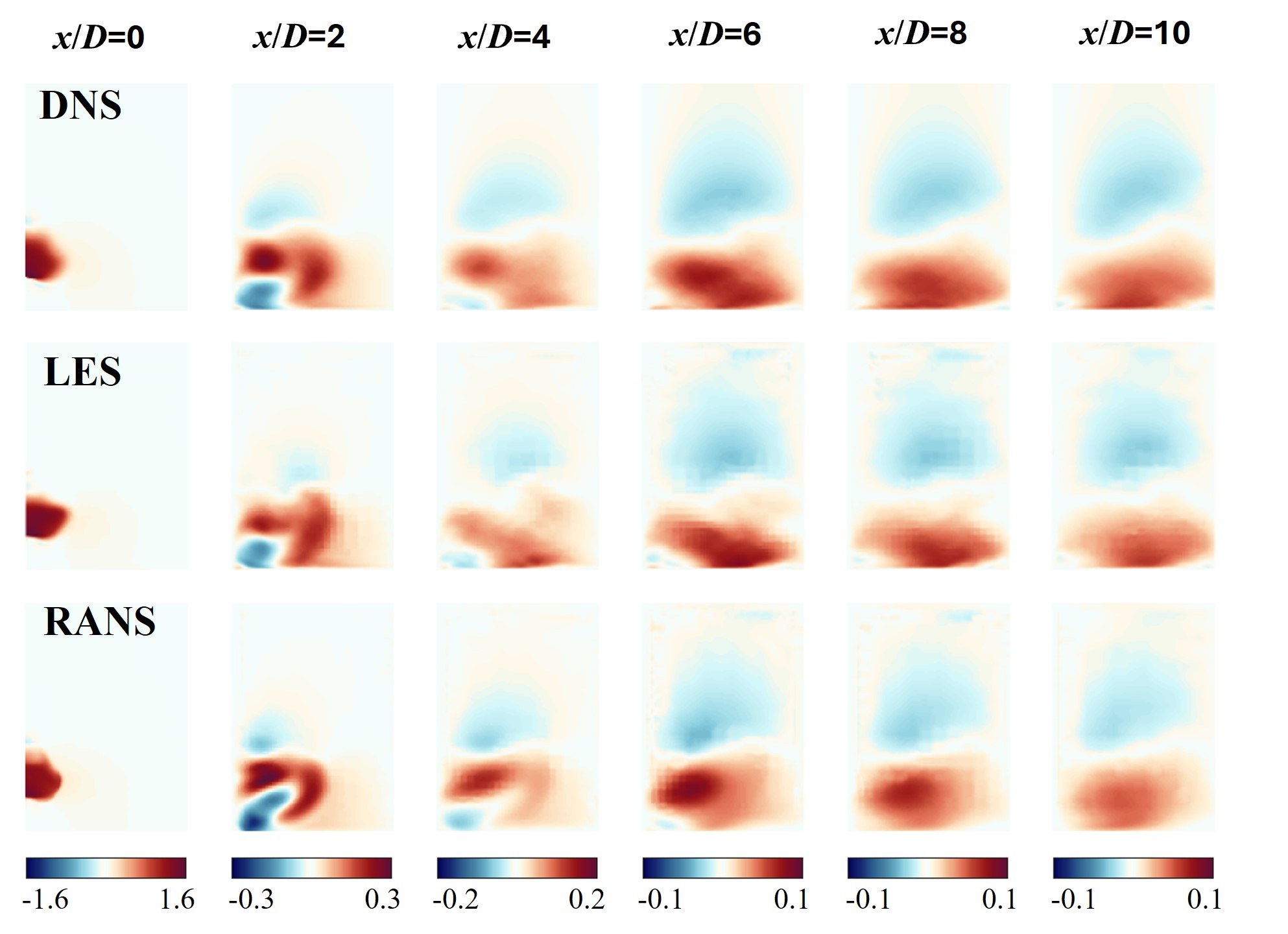}
\caption{Time-averaged velocity component $\overline{v}/U_{jet}$ at different vertical planes with $x/D$=0, 2, 4, 6, 8, and 10, predicted by DNS, LES, and RANS.}\label{fig_vertical_v}
\end{figure}

\begin{figure}[tbp]%% placement specifier
\centering%% For centre alignment of image.
\includegraphics[width=0.8\linewidth]{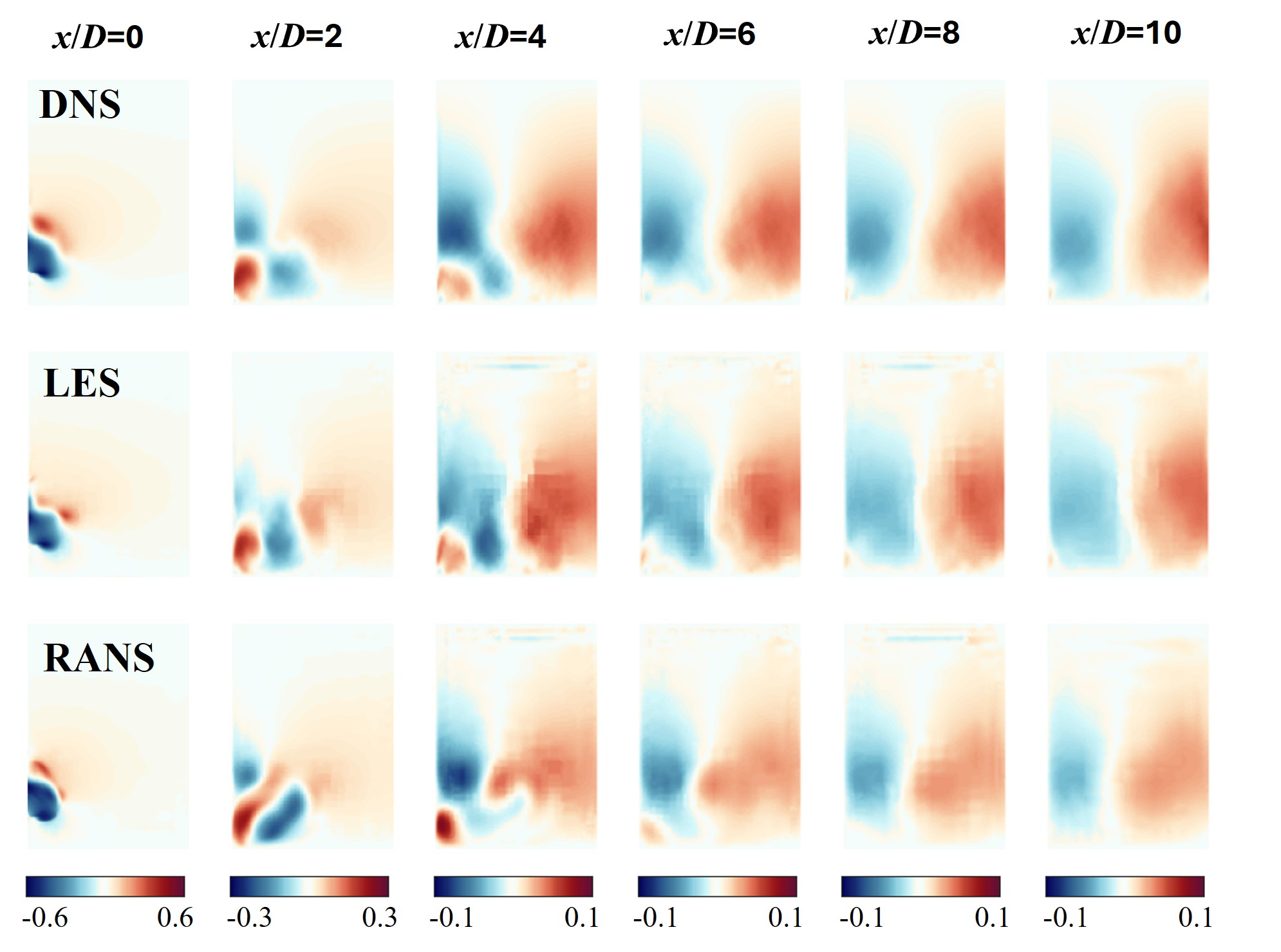}
\caption{Time-averaged velocity component $\overline{w}/U_{jet}$ at different vertical planes with $x/D$=0, 2, 4, 6, 8, and 10, predicted by DNS, LES, and RANS.}\label{fig_vertical_w}
\end{figure}

\begin{figure}[tbp]%% placement specifier
\centering%% For centre alignment of image.
\includegraphics[width=0.8\linewidth]{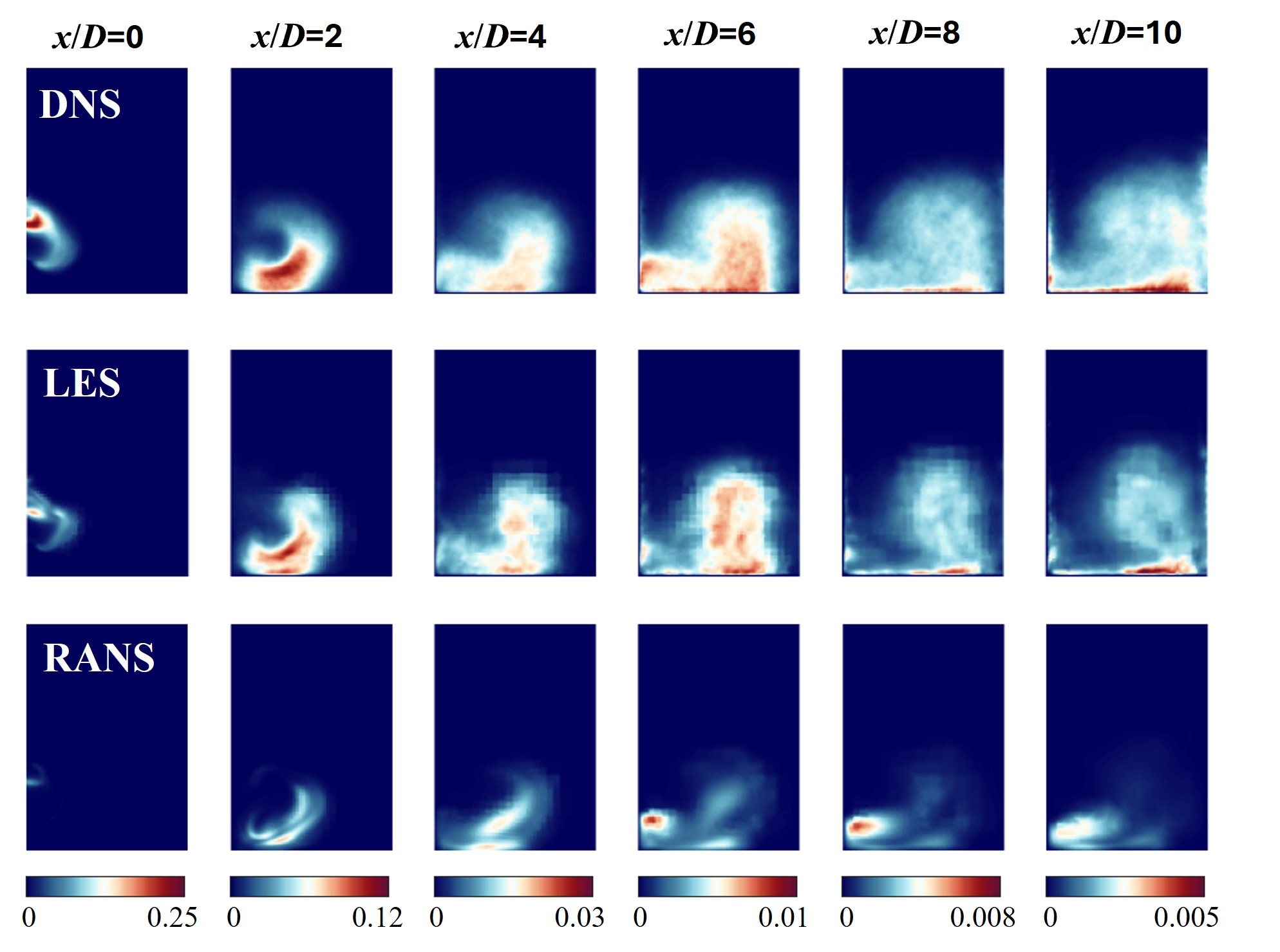}
\caption{Time-averaged Reynolds stress component $\overline{u'u'}/U_{jet}^2$ at different vertical planes with $x/D$=0, 2, 4, 6, 8, and 10, predicted by DNS, LES, and RANS.}\label{fig_vertical_uu}
\end{figure}

\begin{figure}[tbp]%% placement specifier
\centering%% For centre alignment of image.
\includegraphics[width=0.8\linewidth]{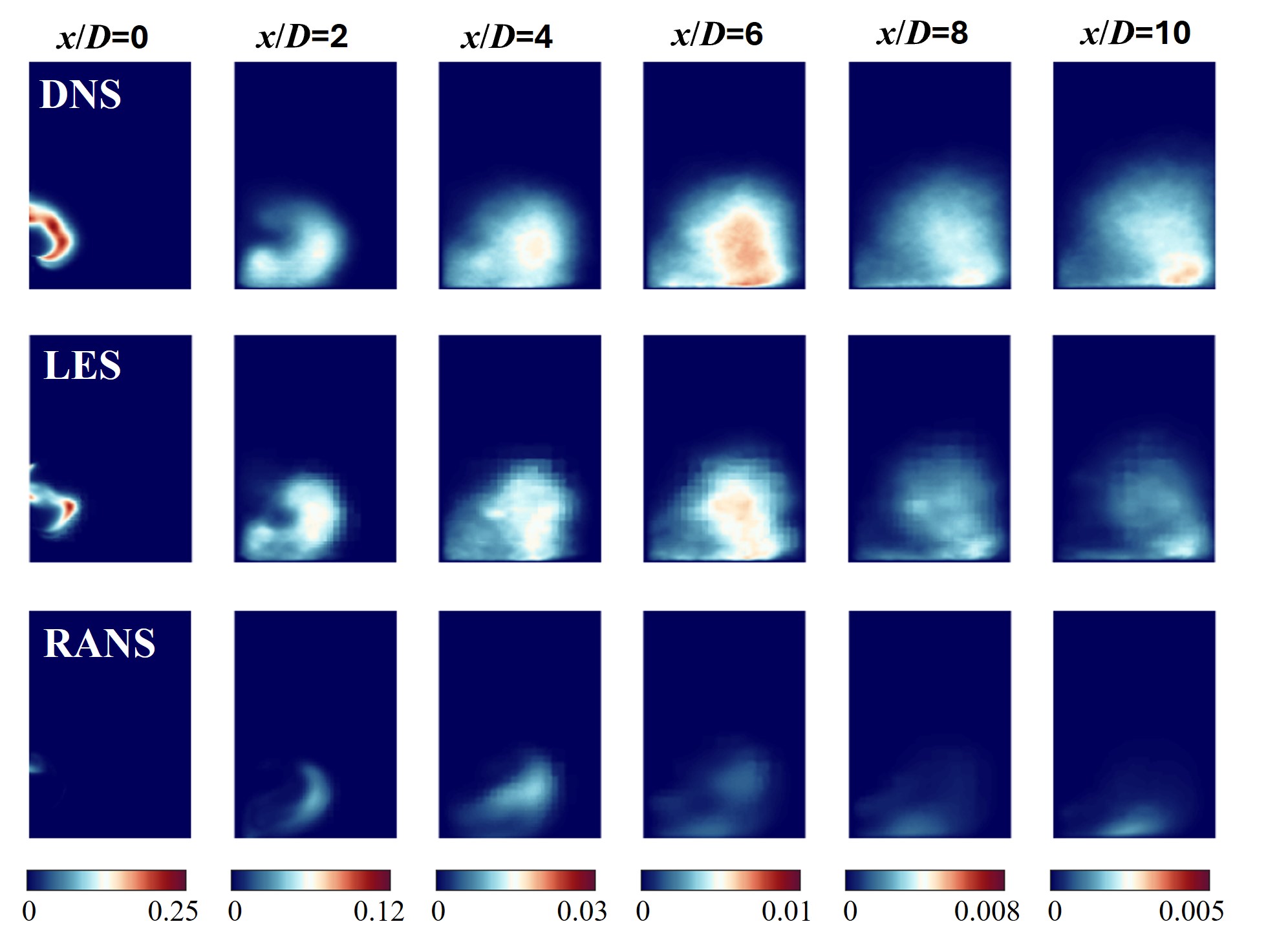}
\caption{Time-averaged Reynolds stress component $\overline{v'v'}/U_{jet}^2$ at different vertical planes with $x/D$=0, 2, 4, 6, 8, and 10, predicted by DNS, LES, and RANS.}\label{fig_vertical_vv}
\end{figure}

\begin{figure}[tbp]%% placement specifier
\centering%% For centre alignment of image.
\includegraphics[width=0.8\linewidth]{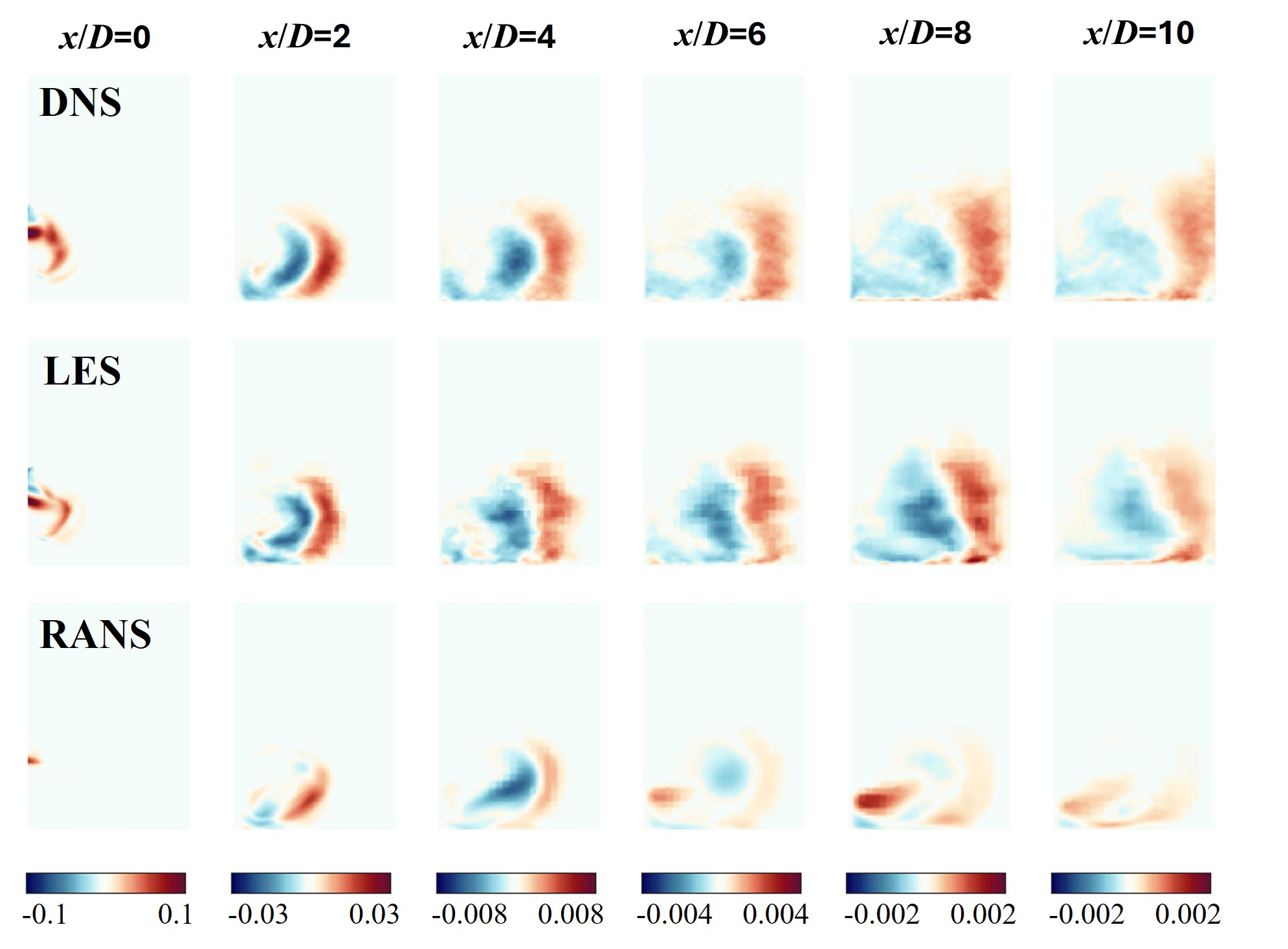}
\caption{Time-averaged Reynolds stress component $\overline{u'v'}/U_{jet}^2$ at different vertical planes with $x/D$=0, 2, 4, 6, 8, and 10, predicted by DNS, LES, and RANS.}\label{fig_vertical_uv}
\end{figure}

Figures~\ref{fig_vertical_u}-\ref{fig_vertical_w} further show the time-averaged velocity components ($\overline{u}$, $\overline{v}$, and $\overline{w}$) at different vertical planes from jet exit ($x/D$=0) to the crossflow outlet ($x/D$=10), predicted by DNS, LES, and RANS. It is found that LES has a much better agreement with DNS than RANS. Specifically, in DNS and LES, the initial high $\overline{u}$ jet quickly evolves into a ring-shaped cloud and then slowly transitions into a round-shaped diffused cloud. In the RANS results, however, $\overline{u}$ decays along the streamwise direction in much slower rate, resulting in a band-shaped high $\overline{u}$ region near the bottom wall at downstream locations ($x/D \geq$2). Besides, considerable differences in the distributions of $\overline{v}$ and $\overline{w}$ are also observed near the bottom left corner of the planes with $x/D$=4 and 6. However, it is worth mentioning that despite the differences observed above,  the velocity components predicted by RANS are generally in a good agreement with those from DNS, at least qualitatively, especially compared with the predictions for Reynolds stress components, which are shown in Figs. \ref{fig_vertical_uu}-\ref{fig_vertical_uv}.

Figures \ref{fig_vertical_uu}-\ref{fig_vertical_uv} show the time-averaged Reynolds stress components ($\overline{u'u'}$, $\overline{v'v'}$, and $\overline{u'v'}$) at different vertical planes. Similar to previous results, LES still shows comparable predictions to DNS regarding Reynolds stress components. However, dramatic differences are seen between the results from RANS and DNS. Specifically, RANS significantly under-predicts all the Reynolds stress components at all the streamwise locations. The same phenomena are also observed for the distributions of $\overline{w'w'}$, $\overline{v'w'}$, and $\overline{w'u'}$ which are not plotted here for brevity. The interested readers may refer to Supplementary Materials for detailed results. 
 
To summarize, the flow field results indicate that LES performs well in predicting both the mean flow velocity and the Reynolds stress compared to DNS data. In contrast, RANS does not predict the Reynolds stress well while its prediction of the mean flow velocity is somewhat satisfactory. 

\subsubsection{H$_2$ distribution predictions}
\begin{figure}[tbp]%% placement specifier
\centering%% For centre alignment of image.
\includegraphics[width=1.0\linewidth]{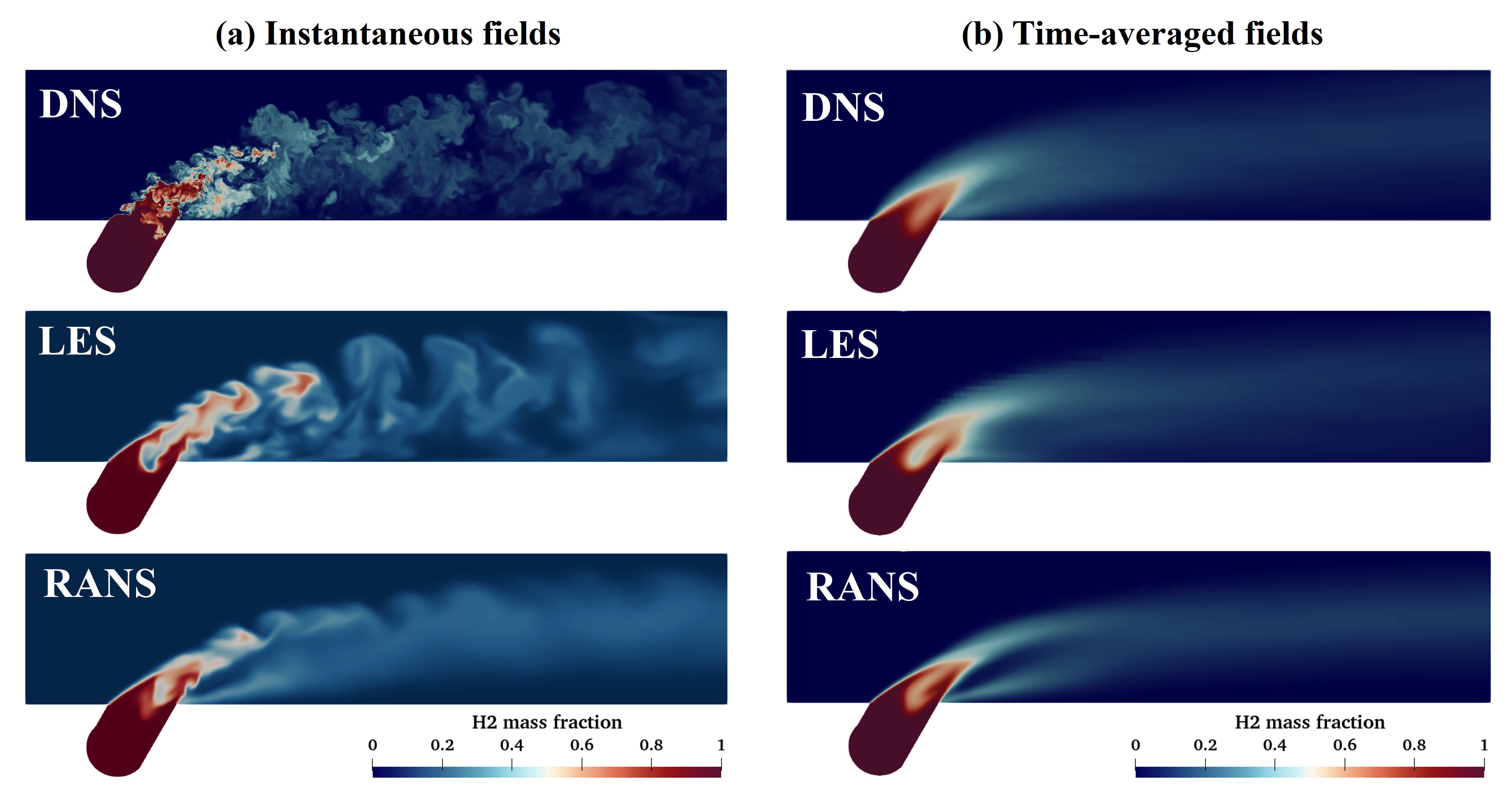}
\caption{Instantaneous (left) and time-averaged (right) H$_2$ mass fraction distributions at the horizontal plane with $z/D$=1, predicted by DNS, LES, and RANS.}\label{fig_yh2_horizontal}
\end{figure}

The predictions of H$_2$ mixing process by DNS, LES, and RANS are further examined. Figure~\ref{fig_yh2_horizontal} shows the instantaneous and time-averaged distributions of H$_2$ mass fraction in the horizontal plane with $z/D$=1. Similar to Fig.~\ref{fig_flow_horizontal}, DNS shows fine mixing structures in the instantaneous H$_2$ mass fraction field, which is captured by LES to some extent. For the time-averaged field, it can be clearly seen that the LES prediction agrees better with DNS data, compared to RANS, especially at the periphery of the jet and downstream regions. Especially, RANS predicts a clear separation between the primary H$_2$ jet and the secondary branch near the bottom wall, resulting in a low-H$_2$ region on the leeward side of the jet. Such low-H$_2$ region is not observed in DNS or LES, suggesting that the mixing between these two H$_2$ branches is under-predicted by the current RANS model .

\begin{figure}[tbp]%% placement specifier
\centering%% For centre alignment of image.
\includegraphics[width=0.9\linewidth]{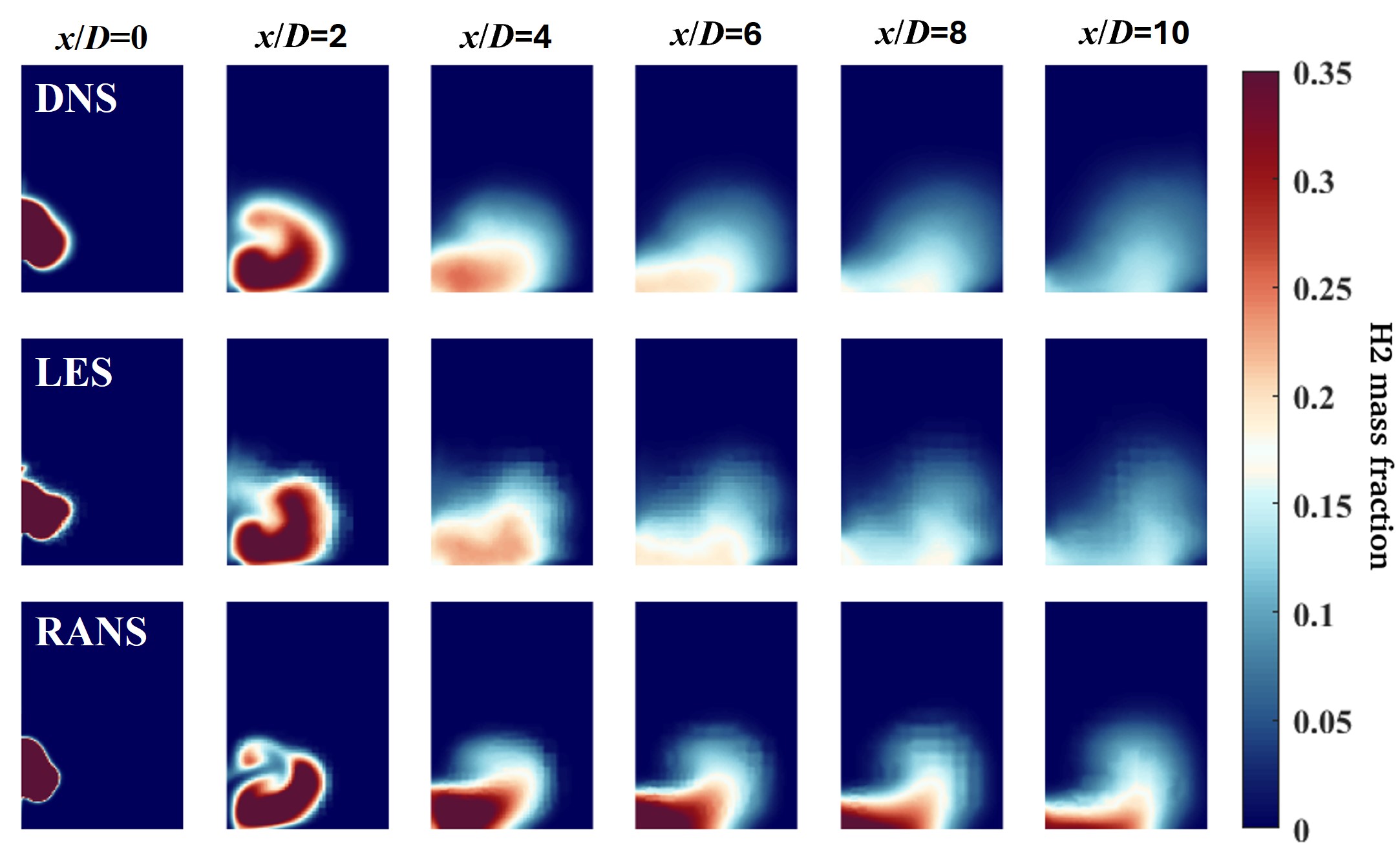}
\caption{Time-averaged H$_2$ mass fraction distributions at different vertical plane with $x/D$=0, 2, 4, 6, 8, and 10, predicted by DNS, LES, and RANS.}\label{fig_vertical}
\end{figure}

To comprehensively evaluate the performance of LES and RANS models in predicting H$_2$ mixing in JICF, Figure~\ref{fig_vertical} compares the time-averaged H$_2$ mass fraction distributions at different vertical planes spanning from $x/D$=0 to 10. At $x/D$=2 near the jet exit, a kidney-shaped H$_2$ distribution is observed in DNS, due to the presence of the counter-rotating vortex pair. However, since the H$_2$ jet center is closer to the bottom wall than the top wall and there is an overall downward velocity due to the vertical section of the injection tube, the mixing process in the lower part is more suppressed compared to the upper part. Hence, the lower half of the kidney-shaped profile is larger and features higher H$_2$ concentration compared with the upper half. As the mixture moves downstream, H$_2$ continues to mix with surrounding air. Consequently, the H$_2$ cloud grows while the peak value of H$_2$ mass fraction keeps decreasing. LES results again show excellent agreement with DNS. Similar trend is observed in RANS as well, however, compared with DNS and LES, RANS predicts a higher concentration region at the bottom wall and shows less mixing with the air in the top part of the channel, consistent with the velocity field. 

\begin{figure}[tbp]%% placement specifier
\centering%% For centre alignment of image.
\includegraphics[width=0.9\linewidth]{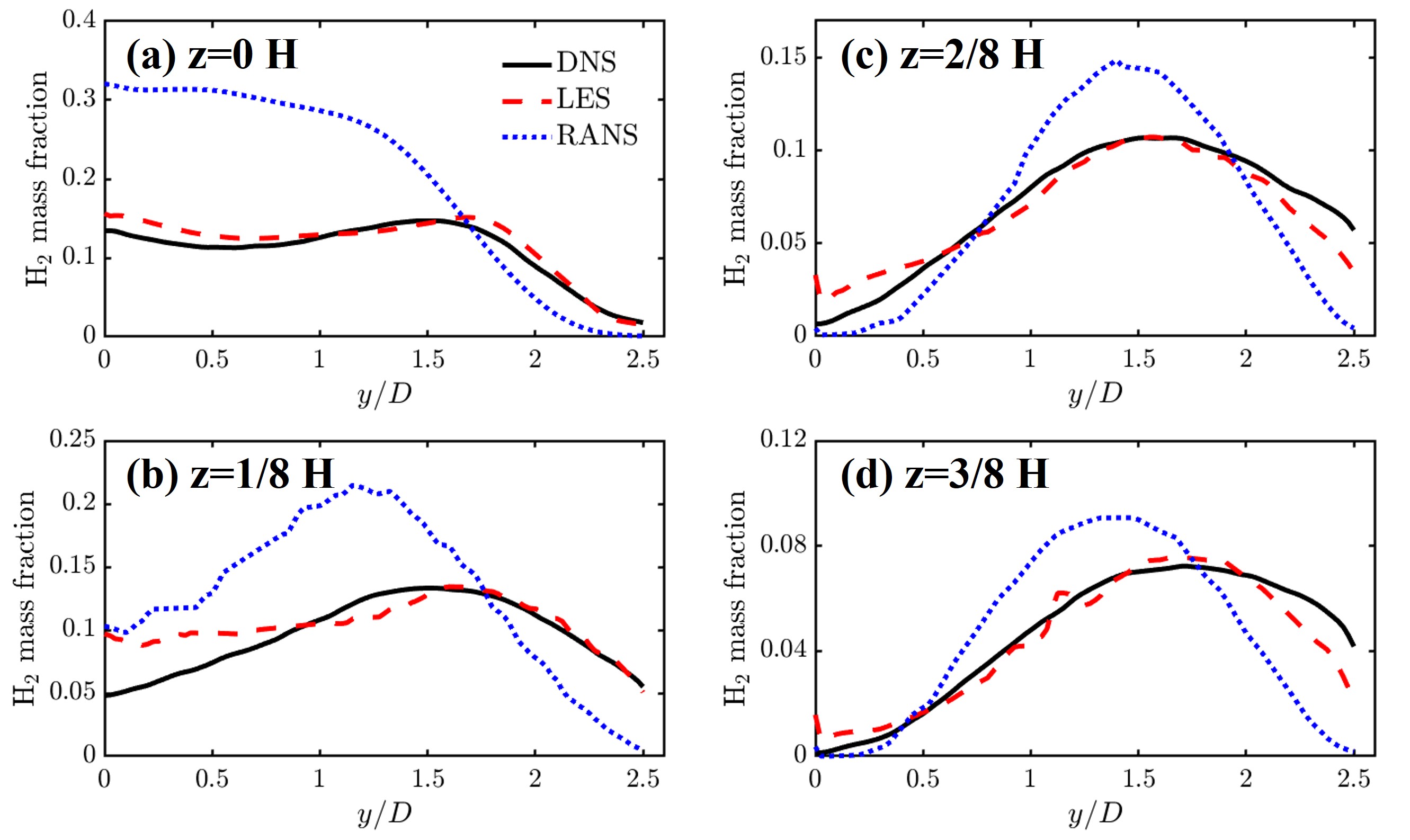}
\caption{Time-averaged H$_2$ mass fraction profiles along the transverse ($y$) direction at different heights (i.e., $z/H$=0, 1/8, 2/8, and 3/8 where $H$=3.6$D$ is the dimension of the channel in spanwise ($z$) direction) in the outlet vertical plane ($x/D$=10). }\label{fig_lineplots}
\end{figure}

In particular, at the outlet plane ($x/D$=10, the right-most column in Fig.~\ref{fig_vertical}), the kidney-shaped profile has almost completely vanished in DNS and LES results with H$_2$ occupying the majority of the lower half corner of the channel, resulting in a fair homogeneous H$_2$ in that region. On the contrary, RANS under-predicts H$_2$ mixing as indicated by the smaller H$_2$ cloud and higher peak value of H$_2$ mass fraction. This is further quantified in Fig.~\ref{fig_lineplots}, where the time-averaged H$_2$ mass fraction profiles along the transverse ($y$) direction at different heights of the outlet vertical plane are plotted. Compared with DNS and LES, at $z$=0 H height, RANS predicts a much higher H$_2$ mass fraction in the region near the side wall (from which the jet is injected into the crossflow) but much lower H$_2$ mass fraction in the region far away from jet. At higher locations, such over-prediction from RANS shifted to the center region (0.5$<y/D<$2). In contrast, the agreement between DNS and LES is excellent, with only marginal deviation near the wall. 

\begin{figure}[tbp]%% placement specifier
\centering%% For centre alignment of image.
\includegraphics[width=0.9\linewidth]{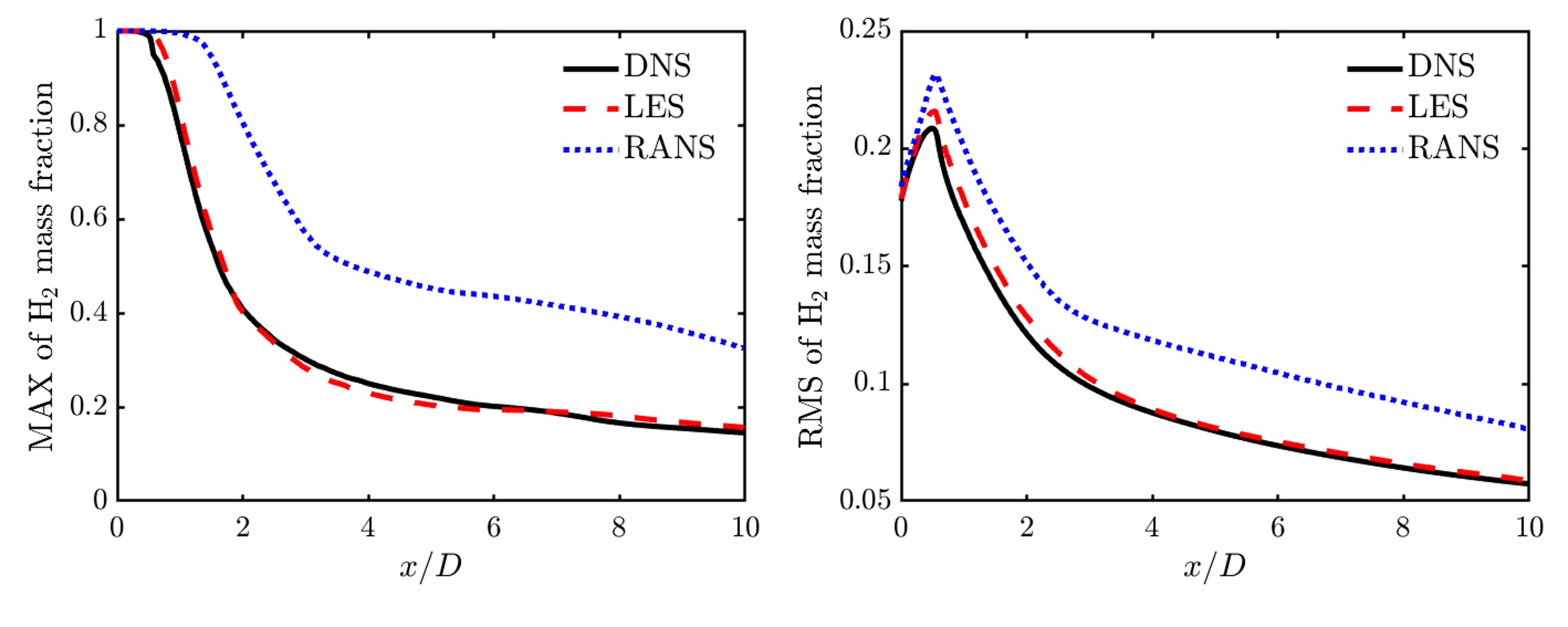}
\caption{The maximum (left) and root-mean-square (right) of time-averaged H$_2$ mass fraction at different vertical planes. }\label{fig_lineplots_streamwise}
\end{figure}

Figure~\ref{fig_lineplots_streamwise} further quantifies the maximum (MAX) and root-mean-square (RMS) of time-averaged H$_2$ mass fraction at different vertical planes predicted by DNS, LES, and RANS. It is seen that LES predictions are in a very good agreement with DNS for both MAX and RMS along the streamwise direction. Although similar trend along the streamwise direction is captured, RANS significantly over-predicts the MAX and RMS of H$_2$ mass fraction at each downstream location, indicating H$_2$ is less mixed with air crossflow in RANS compared with LES and DNS. 

The above results clearly demonstrate that the current RANS model lacks the capability to satisfactorily predict the mixing process between H$_2$ jet and air crossflow. Considering the significant computational cost benefit of RANS compared to LES and DNS, it is of great  practical relevance to gain deeper understanding of the root causes for the deficiencies in the RANS mixing model.

\subsection{Assessment of turbulent species flux models in RANS}
To capture turbulent mixing in RANS, the turbulent species flux model is required. In this section, the commonly used turbulent species flux model in RANS, i.e., the GDH model, will be thoroughly examined based on DNS data. Specifically, the ground-truth values of all turbulent transport properties that are involved in turbulent species flux model will be directly extracted based on the DNS data and compared with those used in RANS. The goal is to identify the origins for the large deficiency of GDH model observed in previous section. Then, the performance of GDH model on characterizing the anisotropy of turbulent diffusivity is evaluated  to provide insights for the further model improvement.

\subsubsection{Turbulent transport properties}
In RANS simulation for multi-component flows, the turbulent species flux term $\widetilde{Y_k^{''}u_j^{''}}$ is usually modeled based on GDH as follows:
\begin{equation}
\widetilde{Y_k^{''}u_i^{''}} = - D_t \frac{\partial \widetilde{Y_k}}{\partial x_i}
\label{eq_gdh}
\end{equation}
where $Y_k$ is the mass fraction of species $k$, $u_i$ the flow velocity in $i$ direction, and $D_t$ the turbulent diffusivity. Note that unlike Eq.~\ref{eq_gdh_intro} where time-averaging is employed, Favre-averaging is adopted here since there is substantial density variation in the current H$_2$ JICF system. Therefore, $\widetilde{f}$ denotes Favre-averaging of quantity $f$, and $f^{''}$=$f$-$\widetilde{f}$. Typically, the turbulent diffusivity $D_t$ is assumed to be correlated with the turbulent kinematic viscosity $\nu_t$ through a turbulent Schmidt number $Sc_t$ (see Eq. \ref{eq_Dt_intro}). A default value of $Sc_t$=0.78 is typically used in engine CFD applications. 

The turbulent kinematic viscosity $\nu_t$ is introduced for the closure of the Reynolds stress term in the momentum equation as follows:
\begin{equation}
-\widetilde{u_i^{''}u_j^{''}}=2\nu_t S_{ij}-\frac{2}{3}\delta_{ij}(k+\nu_t \frac{\partial \widetilde{u_i}}{\partial x_i})
\label{eq_reynolds_stress}
\end{equation}
where $S_{ij}$=$\frac{1}{2}(\frac{\partial \widetilde{u_i}}{\partial x_j}+\frac{\partial \widetilde{u_j}}{\partial x_i})$ is strain-rate tensor, and $k$=$\frac{1}{2}\widetilde{u_i^{''}u_i^{''}}$ is the turbulent kinetic energy. 

In Eqs. (\ref{eq_gdh}) and (\ref{eq_reynolds_stress}), all terms except the turbulent transport properties ($D_t$ and $\nu_t$) can be directly obtained from the DNS data by the time-averaging and Favre-averaging operations. As such, the representative values of $D_t$ and $\nu_t$ can be inferred using a ``least-squares sense" optimization method based on Eqs. (\ref{eq_gdh}) and (\ref{eq_reynolds_stress}), as follows:
\begin{equation}
D_t = \frac{-(\frac{\partial \widetilde{Y_k}}{\partial x_j})(\widetilde{Y_k^{''}u_j^{''}})}{(\frac{\partial \widetilde{Y_k}}{\partial x_j})(\frac{\partial \widetilde{Y_k}}{\partial x_j})}
\label{eq_Dt_derive}
\end{equation}

\begin{equation}
v_t = \frac{(2S_{ij}-\frac{2}{3}\delta_{ij}\frac{\partial \widetilde{u_i}}{\partial x_i})(-\widetilde{u_i^{''}u_j^{''}}+\frac{2}{3}\delta_{ij}k)}{(2S_{ij}-\frac{2}{3}\delta_{ij}\frac{\partial \widetilde{u_i}}{\partial x_i})(2S_{ij}-\frac{2}{3}\delta_{ij}\frac{\partial \widetilde{u_i}}{\partial x_i})}
\label{eq_nut_derive}
\end{equation}
Then, the turbulent Schmidt number $Sc_t$ can obtained as:
\begin{equation}
Sc_t=\nu_t / D_t
\label{eq_sct_derive}
\end{equation}

\begin{figure}[tbp]%% placement specifier
\centering%% For centre alignment of image.
\includegraphics[width=0.8\linewidth]{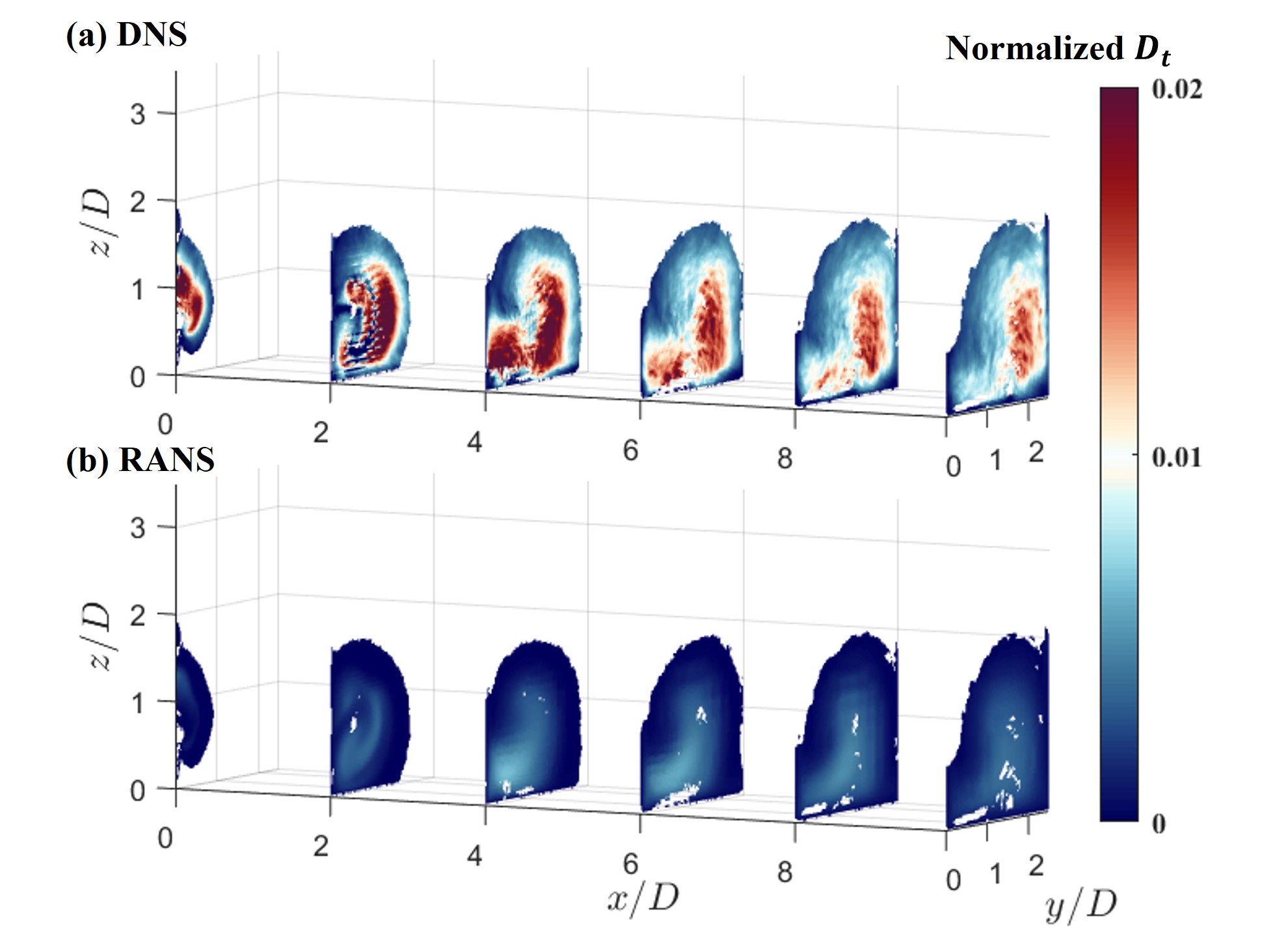}
\caption{Normalized turbulent diffusivity $D_t/(U_{jet}D)$ from DNS and RANS at different vertical planes. Regions with $|\nabla \widetilde{Y_{H_2}}|<$0.05 are blanked.}\label{fig_diffusivity}
\end{figure}

Figure~\ref{fig_diffusivity} compares the turbulent diffusivity $D_t$ between DNS and RANS in vertical planes. Note that in the region with small gradient of H$_2$ mass fraction, $D_t$ is ill-defined, and thus regions with  $|\nabla \widetilde{Y_k}|<$0.05 are blanked. DNS results show a high diffusivity near the jet core throughout the domain and relatively a low diffusivity in the interface between H$_2$ and surrounding air. For RANS, although it qualitatively predicts the distribution of $D_t$, the magnitude of $D_t$ is significantly under-predicted throughout the domain. This clearly explains the under-prediction of H$_2$ mixing process by RANS as shown in previous sections.

\begin{figure}[tbp]%% placement specifier
\centering%% For centre alignment of image.
\includegraphics[width=0.8\linewidth]{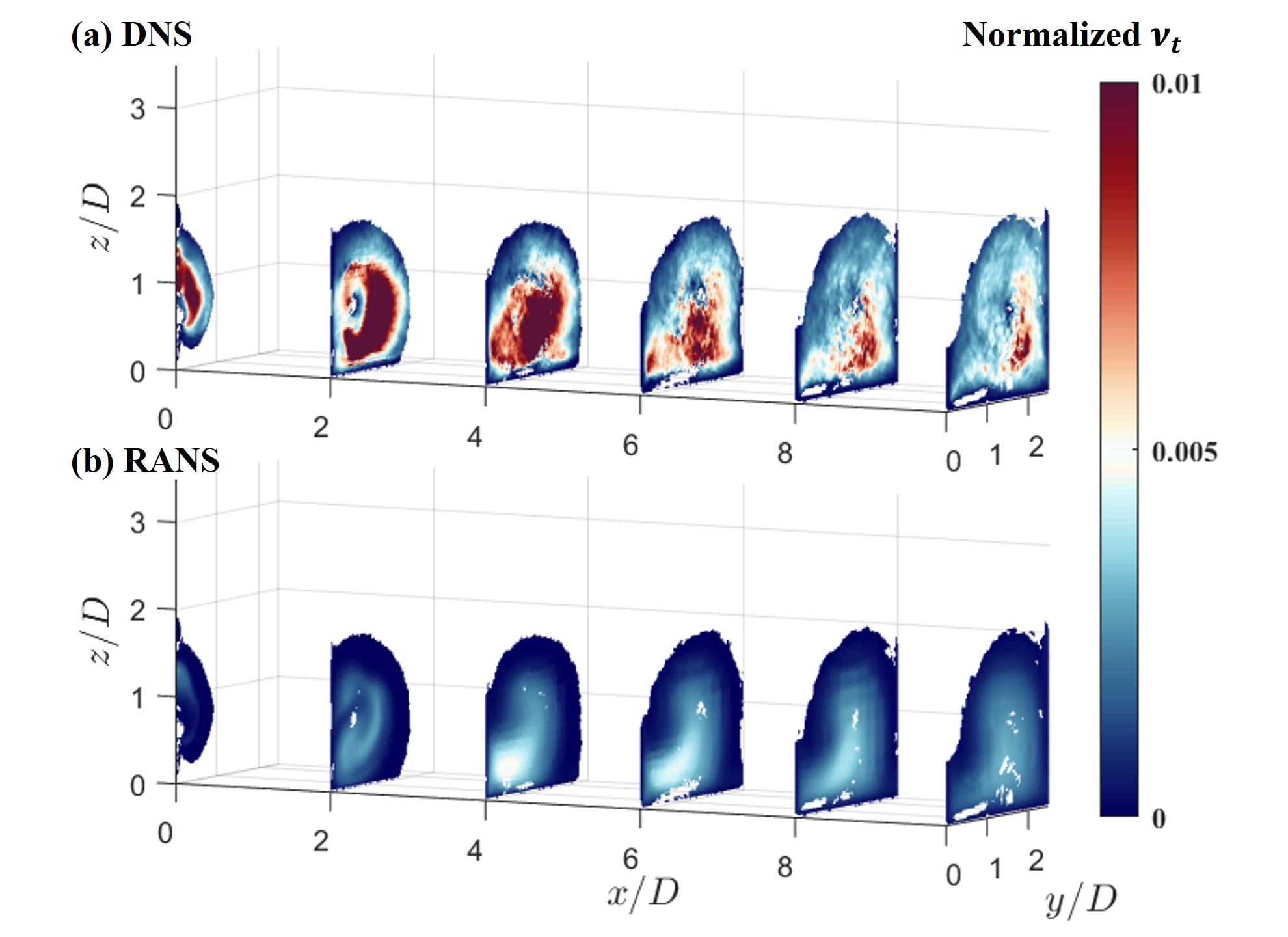}
\caption{Normalized turbulent viscosity $\nu_t/(U_{jet}D)$ from DNS and RANS at different vertical planes. Regions with $|\nabla \widetilde{Y_{H_2}}|<$0.05 are blanked.}\label{fig_viscosity}
\end{figure}

\begin{figure}[tbp]%% placement specifier
\centering%% For centre alignment of image.
\includegraphics[width=0.8\linewidth]{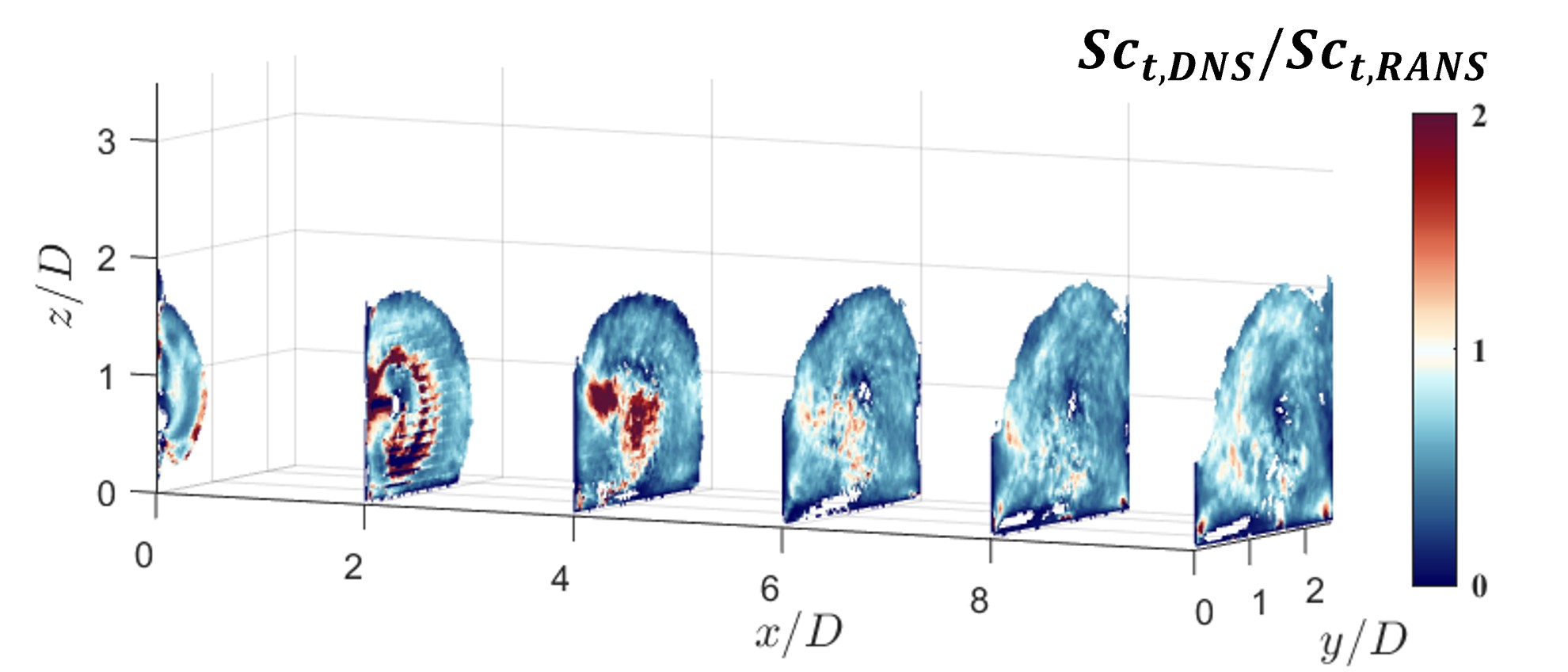}
\caption{Turbulent Schmidt number $Sc_k$ from DNS at different vertical planes. Regions with $|\nabla \widetilde{Y_{H_2}}|<$0.05 are blanked.}\label{fig_sck}
\end{figure}

The value of $D_t$ is determined by both $\nu_t$ and $Sc_t$. In order to reveal the key contributing factor for the difference in $D_t$ between DNS and RANS, Figs. \ref{fig_viscosity} and \ref{fig_sck} show the comparisons between DNS and RANS on $\nu_t$ and $Sc_t$, respectively. Note that for $Sc_t$, only the results from DNS are plotted since $Sc_t$ is a constant in RANS ($Sc_{t,RANS}$=0.78). Figure~\ref{fig_viscosity} shows similar phenomenon as observed in Fig.~\ref{fig_diffusivity}. That is, the magnitude of $\nu_t$ is significantly under-predicted by RANS, although it predicts the shape of the overall distribution, i.e., high $\nu_t$ near the jet core and low $\nu_t$ along the interface. Figure~\ref{fig_sck} plots the ratio of $Sc_t$ between DNS and RANS (i.e., $Sc_{t,DNS}/Sc_{t,RANS}$). It is seen that $Sc_{t,DNS}/Sc_{t,RANS}$ varies considerably throughout the domain, with most regions having $Sc_{t,DNS}/Sc_{t,RANS}<$ 1 and some small regions near the jet exit with $Sc_{t,DNS}/Sc_{t,RANS}>$2. Hence, the under-prediction of $D_t$ in RANS is largely attributed to the inaccuracies in the predictions of both $\nu_t$ and $Sc_t$. 

\begin{figure}[tbp]%% placement specifier
\centering%% For centre alignment of image.
\includegraphics[width=1.0\linewidth]{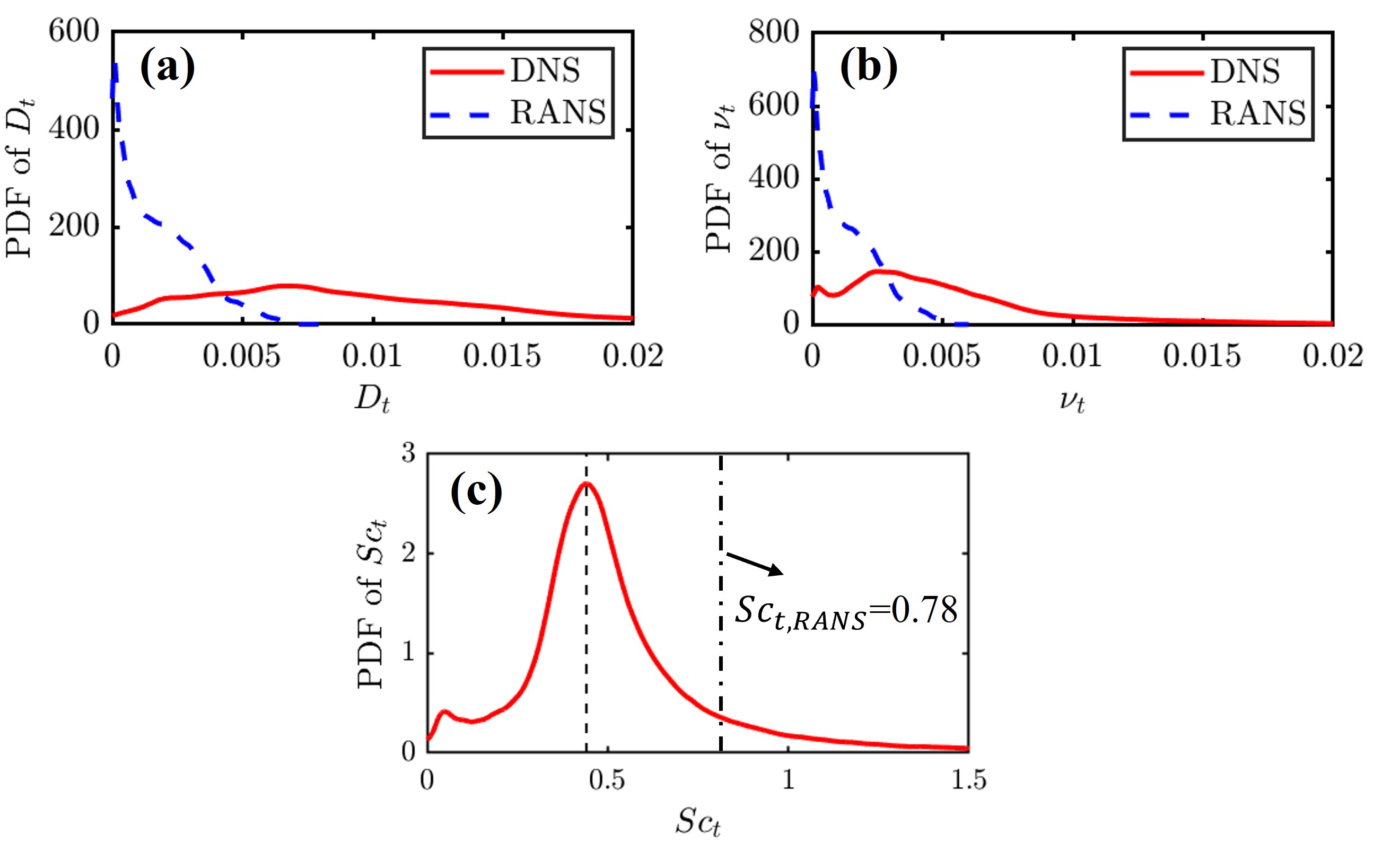}
\caption{Comparison for the PDFs of normalized $D_t$, normalized $v_t$, and $Sc_k$ between DNS and RANS. For $Sc_k$, only DNS results are plotted.The data is collected from the entire domain with $|\nabla \widetilde{Y_k}|>$0.05.}\label{fig_pdf_properties}
\end{figure}

To quantitatively illustrate the difference in $D_t$, $\nu_t$, and $Sc_t$ over the entire domain, Fig.~\ref{fig_pdf_properties} further show the PDFs of normalized $D_t$, normalized $v_t$, and $Sc_k$, which are collected from the 3D domain. Note that the regions with $|\nabla \widetilde{Y_k}|<$0.05 are excluded to rule out nonphysical values. It is seen that the PDFs of $D_t$ and $\nu_t$ derived from DNS span a wider range towards higher values than those from RANS. In the meantime, the peak of $Sc_t$ PDF obtained from DNS is considerably smaller than the default value used in the current RANS model ($Sc_{t,RANS}$=0.78). These are consistent with observations in Figs. \ref{fig_diffusivity}-\ref{fig_sck}. It is also worth mentioning that the most probable value of $Sc_t$ observed in the present study is within the range reported in the literature \cite{2013Ivanova,2016Ling,2017Ryan} where the optimal value of $Sc_t$ was derived based on LES data rather than DNS.

To summarize, the above results indicate that the under-prediction of RANS on H$_2$ mixing process are attributed to both the under-prediction of turbulent viscosity $\nu_t$ and the high value of $Sc_t$ used in RANS. The former aspect has been rarely reported in the literature. However, the present study suggests that this should not be neglected in the context of model development for the purpose of improving models for H$_2$ mixing. More detailed investigation on how to improve $\nu_t$ prediction is out of the scope of this work, and merits further investigation in the future. In the next subsection, the analysis will be focused on the anisotropy of $Sc_t$.

\subsubsection{Anisotropy of $Sc_t$}
When adopting the GDH formulation to model turbulent species flux (i.e., Eqs. (\ref{eq_gdh}) and (\ref{eq_Dt_intro})), it is implicitly assumed that the turbulent diffusivity $D_t$ is an isotropic property. To evaluate this assumption, the anisotropic components of $D_t$ in vector form, i.e., $Sc_{t,x}$, $Sc_{t,y}$, and $Sc_{t,z}$, are separately derived from DNS data as:
\begin{equation}
    Sc_{t,x}=-\nu_t \frac{\partial \widetilde{Y_k}}{\partial x}/\widetilde{Y_k^{''}u_x^{''}}, \quad Sc_{t,y}=-\nu_t \frac{\partial \widetilde{Y_k}}{\partial y}/\widetilde{Y_k^{''}u_y^{''}}, \quad Sc_{t,z}=-\nu_t \frac{\partial \widetilde{Y_k}}{\partial z}/\widetilde{Y_k^{''}u_z^{''}}
\label{eq_sct_anisotropic}
\end{equation}

\begin{figure}[tbp]%% placement specifier
\centering%% For centre alignment of image.
\includegraphics[width=0.5\linewidth]{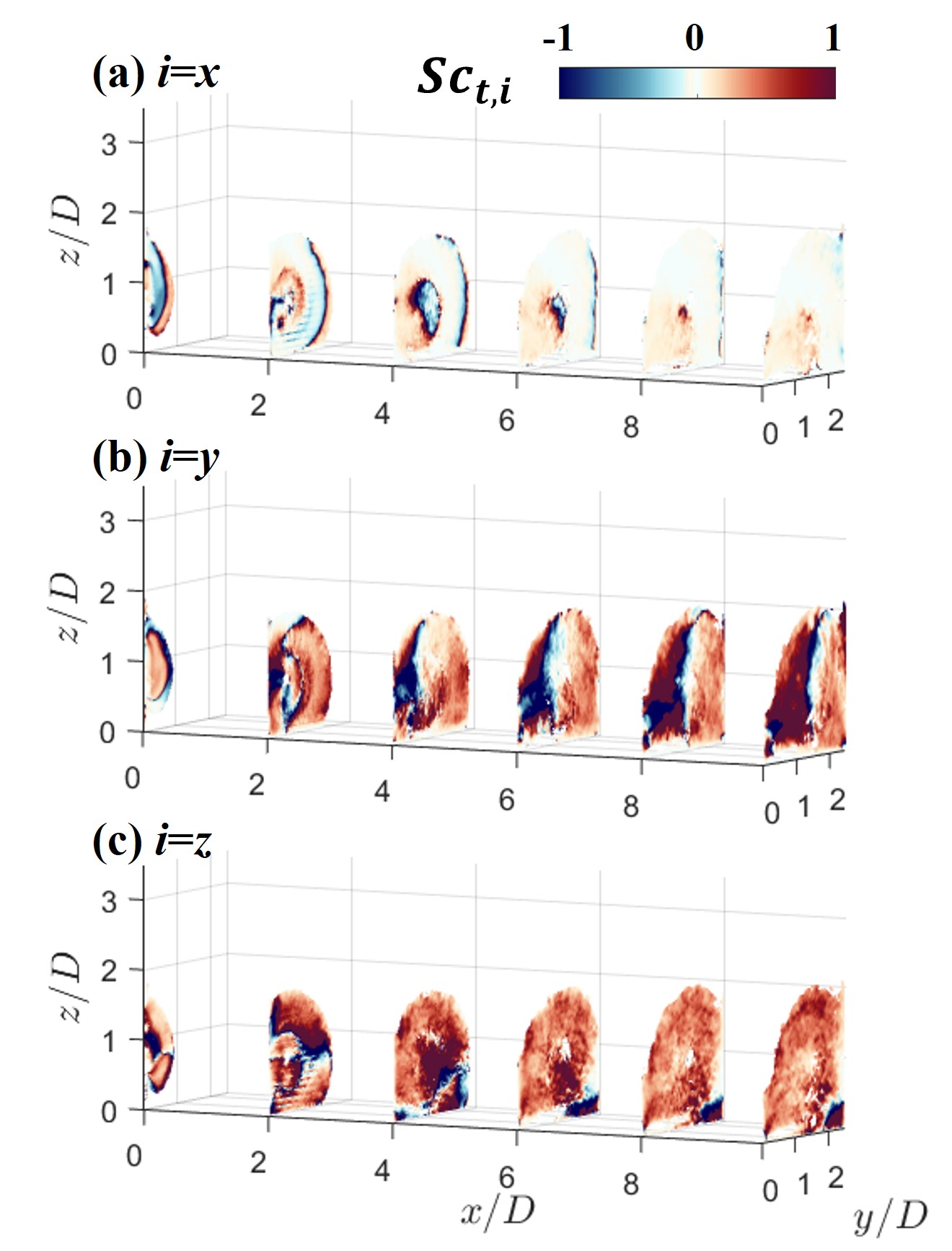}
\caption{Anisotropic components of $Sc_t$ obtained from DNS at different vertical planes. Regions are blanked for $|\nabla \widetilde{Y_{H_2}}|<$0.05.}\label{fig_sck_gdh}
\end{figure}

Figure~\ref{fig_sck_gdh} plots the distributions of these three anisotropic components of $Sc_t$ at different vertical planes. There are some regions with negative $Sc_{t,i}$, indicating the existence of counter-gradient diffusion, but they are limited to small regions and thus do not represent a significant contribution to the overall H$_2$ fluxes. The more important observation is that $Sc_{t,x}$ is much smaller than the other two components. In most regions, $Sc_{t,x}$ is a small value close to zero. By contrast, $Sc_{t,y}$ varies dramatically, especially on the left side of jet ($y/D<$1) where regions with $Sc_{t,y}>$1 are observed, while $Sc_{t,z}$ shows a relatively homogeneous distribution at around 0.5 in regions with $x/D>$4. These results clearly show the deficiency of isotropic $D_t$ in GDH. 

To quantitatively measure the anisotropy of turbulent diffusivity $D_t$, the misalignment angle $\theta$ between the species flux vectors directly obtained from DNS ($\Vec{f}_{dns}$) and predicted by the GDH model ($\Vec{f}_{model}$, the right hand side of Eqs.~(\ref{eq_gdh})) is calculated as:
\begin{equation}
\theta = arccos(\frac{|\Vec{f}_{dns}\cdot \Vec{f}_{model}|}{|\Vec{f}_{dns}||\Vec{f}_{model}|})
\end{equation}
The analysis based on this quantity has the benefit of being independent of the choices of the model constants used since they only act as scaling factors and will not influence $\theta$. A smaller $\theta$ would suggest better alignment between these two flux vectors and thus less anisotropy in $D_t$. 

\begin{figure}[tbp]%% placement specifier
\centering%% For centre alignment of image.
\includegraphics[width=0.6\linewidth]{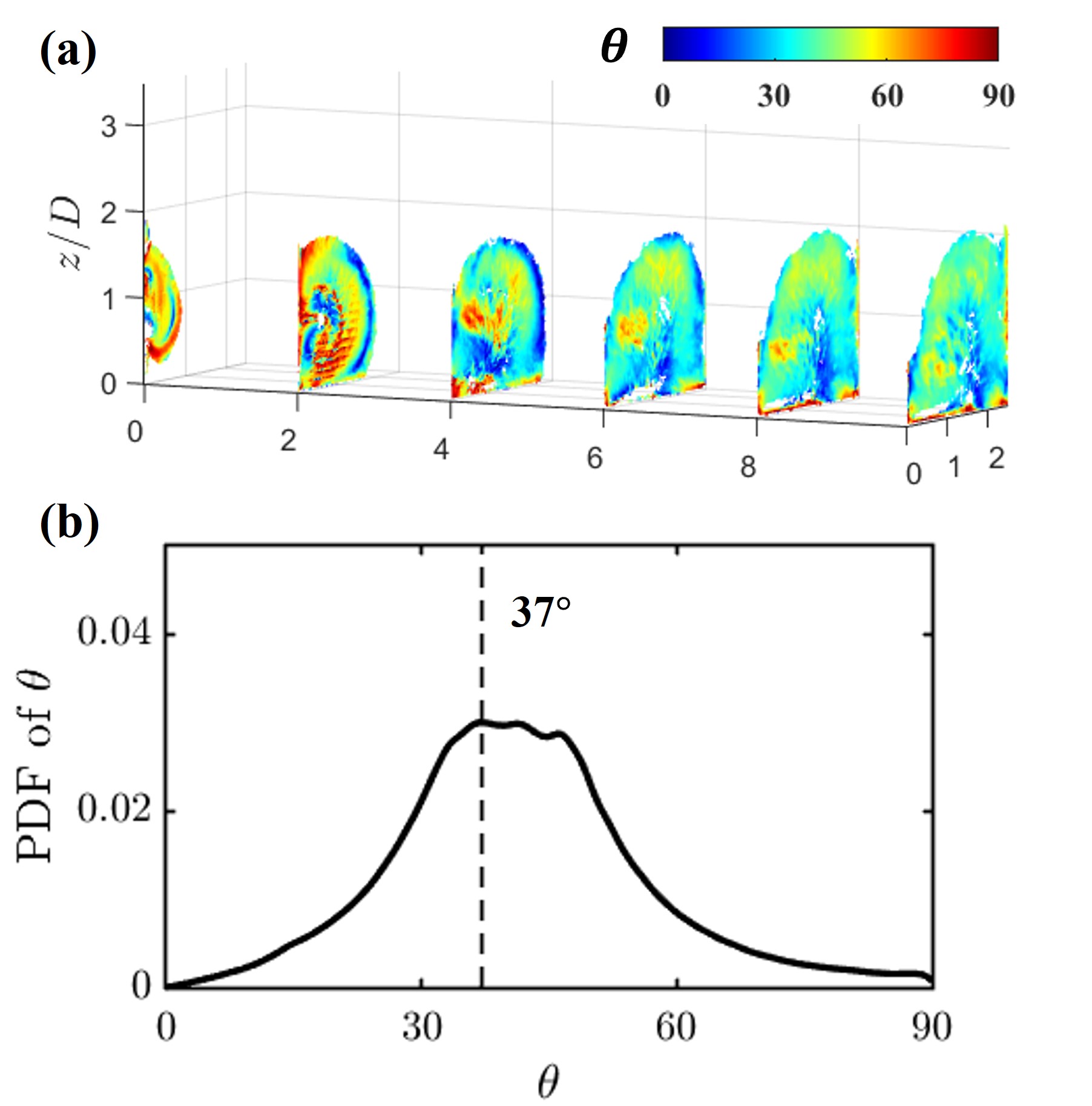}
\caption{(a) The distribution of angle $\theta$ between turbulent species flux vectors from DNS and the GDH model at different vertical planes; (b) PDF of $\theta$ collected from the 3D domain with $|\nabla \widetilde{Y_k}|>$0.05.}\label{fig_theta_gdh}
\end{figure}

Figure~\ref{fig_theta_gdh}(a) shows the distribution of $\theta$ at different vertical planes. It is seen that GDH predicts large values of $\theta$ (near 60$^{\circ}$) in most regions, especially near the jet exit where $\theta$ can be as high as 90$^{\circ}$. Figure~\ref{fig_theta_gdh}(b) plots the PDF of $\theta$ collected from the 3D domain. It is seen that $\theta$ spans a wide range from 0$^{\circ}$ to 90$^{\circ}$ and the PDF peaks at about 37$^{\circ}$, which is considerably larger than 0. These results clearly demonstrate that the anisotropy in $Sc_t$ is not negligible.

\begin{figure}[tbp]%% placement specifier
\centering%% For centre alignment of image.
\includegraphics[width=1.0\linewidth]{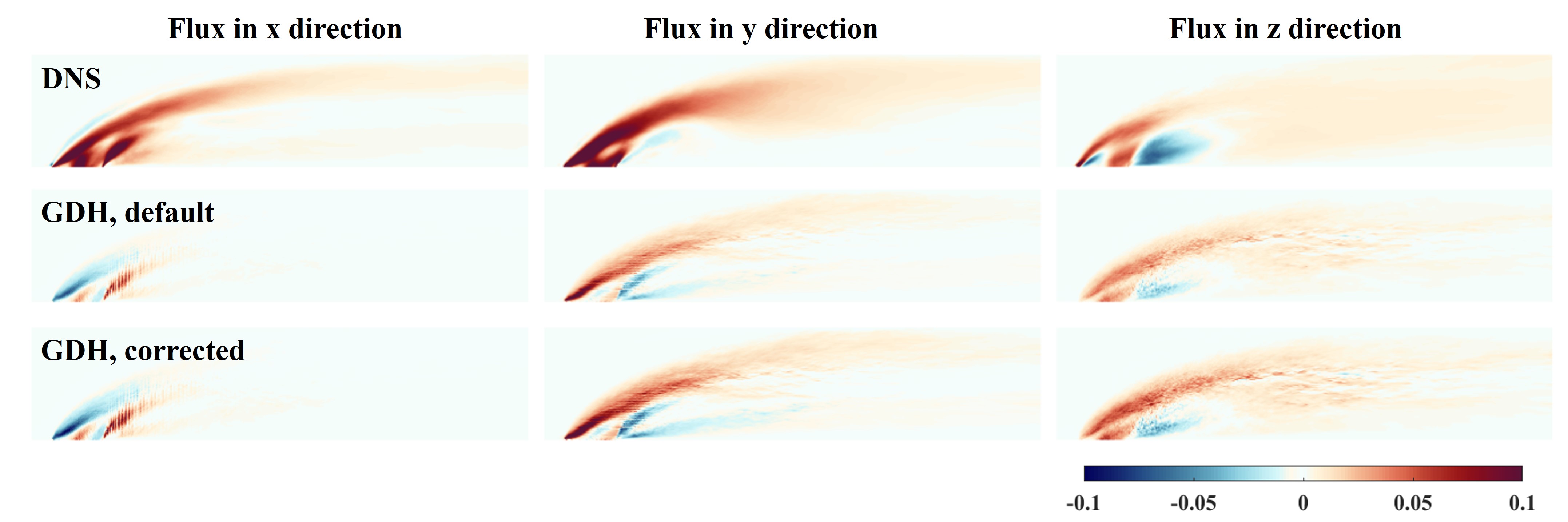}
\caption{Turbulent species fluxes directly obtained from DNS and predicted by the GDH model with two different $Sc_t$ values at the horizontal plane with $z/D$=1.}\label{fig_flux_gdh}
\end{figure}

Finally, in Fig.~\ref{fig_flux_gdh}, the species fluxes predicted from GDH model with two different values of $Sc_t$, one is the default value of 0.78 and the other is 0.44 which is the most portable value derived from DNS data (see Fig.~\ref{fig_pdf_properties}(c)), are directly compared against DNS results. It is seen that the GDH model with default $Sc_t$ has poor performance on predicting the species fluxes. Specifically, it under-predicts the positive species fluxes in most regions and negative species fluxes in $y$ and $z$ directions in the leeward of jet, while also over-predicts the negative species fluxes in $x$ direction in the windward of the jet. By using a corrected $Sc_t$, the performance of GDH model is improved, however, only to some limited extent, as there are still some regions showing large discrepancies compared with DNS results. For example, the over-prediction for the negative species flux in $x$ direction is amplified by the corrected $Sc_t$. The above results clearly demonstrate the importance of accounting for anisotropy  for turbulent mixing model in the context of H$_2$ jet mixing process. Therefore, in future work, it would be of great interest to evaluate the performance of higher-order models (such as GGDH and HOGGDH) which take the anisotropy of turbulent diffusivity into account.

\section{Conclusions}
In this work, a numerical study that employs DNS, LES, and RANS approaches was conducted for H$_2$ jet in a crossflow based on a geometry representative of key aspects of port fuel injection in a H$_2$-fueled heavy-duty ICE. Based on the DNS data, the performance of LES and RANS on predicting the turbulent flow fields and mixing process in this specific configuration was comprehensively evaluated. It was found that LES has very good performance on predicting not only the mean velocity components but also the Reynolds stress components. In contrast, RANS significantly under-predicts all components of Reynolds stress, although it shows relatively good performance on predicting the mean flow field. Regarding the H$_2$ mixing prediction, LES results also show excellent agreement with DNS results, while RANS still significantly under-predicts the mixing process. 

Then, the underlying reasons for the poor performance of RANS were investigated by extracting turbulent transport properties used in RANS approach from DNS data. Results showed that the turbulent diffusivity predicted by RANS is much smaller than that derived from DNS, which explains the under-prediction of RANS for mixing process. Besides, the under-prediction of turbulent diffusivity is not only attributed to the over-prediction on turbulent Schmidt number, but also the under-prediction on turbulent viscosity. The latter also contributes to the under-prediction of Reynolds stress components observed in RANS. Furthermore, by analyzing the anisotropic components of turbulent Schmidt number and the misalignment angle between turbulent species fluxes directly obtained from DNS and predicted by the mixing model used in RANS, the commonly used assumption of isotropic turbulent diffusivity was demonstrated to be invalid for the present configuration.

The present study provides useful information to improve the turbulent species mixing prediction in RANS framework. Besides, the DNS data presented here can be used for future model development work for both turbulent viscosity and turbulent diffusivity.   

\section*{Acknowledgments}

The submitted manuscript has been created by UChicago Argonne, LLC, Operator of Argonne National Laboratory (“Argonne”). Argonne, a U.S. Department of Energy Office of Science laboratory, is operated under Contract No. DE-AC02-06CH11357. The U.S. Government retains for itself, and others acting on its behalf, a paid-up nonexclusive, irrevocable worldwide license in said article to reproduce, prepare derivative works, distribute copies to the public, and perform publicly and display publicly, by or on behalf of the Government. The Department of Energy will provide public access to these results of federally sponsored research in accordance with the DOE Public Access Plan (http://energy.gov/downloads/doe-public-access-plan). 

This work is performed under a Cooperative Research and Development Agreement (CRADA) between Argonne, Caterpillar Inc., and Convergent Science, Inc. The authors would like to thank Gurpreet Singh and Nick Hansford, respectively Program Manager and Technology Manager at the Vehicle Technologies Office, Energy Efficiency and Renewable Energy, Off-road, Rail, Maritime, and Aviation (ORMA) Program, for supporting this work, and Caterpillar and Convergent Science for co-funding this work. The authors would also like to acknowledge the computing resources on the Bebop and Improv clusters provided by the Laboratory Computing Resource Center (LCRC) at Argonne National Laboratory.

\appendix

\bibliographystyle{elsarticle-num-names} 
\bibliography{H2Mixing}





%%\bibitem[Lamport(1994)]{lamport94}
%%  Leslie Lamport,
%%  \textit{\LaTeX: a document preparation system},
%%  Addison Wesley, Massachusetts,
%%  2nd edition,
%%  1994.

%%\end{thebibliography}
\end{document}